# Physics-Informed Neural Networks in Electromagnetic and Nanophotonic Design

Omar A. M. Abdelraouf,[a,*] Abdulrahman M. A. Ahmed,[b] Emadeldeen Eldele,[c] and Ahmed A. Omar [b,d]

[a] Institute of Materials Research and Engineering, Agency for Science, Technology, and Research (A*STAR), 2 Fusionopolis Way, #08-03, Innovis, Singapore 138634, Singapore.

[b] Department of Electrical Engineering, King Fahd University of Petroleum & Minerals, Dhahran, 31261, Saudi Arabia.

[c] Institute for Infocomm Research, A*STAR, Singapore, 138632, Singapore.

[d] Interdisciplinary Research Center for Communication Systems and Sensing, King Fahd University of Petroleum & Minerals, Dhahran, 31261, Saudi Arabia.

* Corresponding author.  Email address:  Omar_Abdelrahman@imre.a-star.edu.sg

## Abstract

The fusion of artificial intelligence (AI) with physics-guided frameworks has opened transformative avenues for advancing the design and optimization of electromagnetic and nanophotonic systems. Innovations in deep neural networks (DNNs) and physics-informed neural networks (PINNs) now provide robust tools to tackle longstanding challenges in light scattering engineering, meta-optics, and nonlinear photonics. This review outlines recent progress in leveraging these computational methodologies to enhance device performance across domains such as dynamic light modulation, antenna design, and nonlinear optical phenomena. We systematically survey advancements in AI-driven forward and inverse design strategies, which bypass conventional trial-and-error approaches by embedding physical laws directly into optimization workflows. Furthermore, the integration of AI accelerates electromagnetic simulations and enables precise modelling of complex optical effects, including topological photonic states and nonlinear interactions. A comparative evaluation of algorithmic frameworks highlights their strengths in balancing computational efficiency, multi-objective optimization, and fabrication feasibility. Challenges such as limited interpretability of AI models and data scarcity for unconventional optical modes are critically addressed. Finally, we emphasize future opportunities in scalable multi-physics modelling, adaptive architectures, and practical deployment of AI-optimized photonic devices. This work underscores the pivotal role of AI in transcending traditional design limitations, thereby propelling the development of next-generation photonic technologies with unprecedented functionality and efficiency.

# 1. Introduction

## 1.1. Overview

Recent studies in nanophotonics and electromagnetism have utilized the advantage of DNNs through improving the design, optimization, and adaptive control of photonic and electromagnetic properties. For example, in the case of antenna design and frequency-selective surfaces (FSS), DNNs accelerate inverse design by directly linking target electromagnetic properties, such as broadband operation or reconfigurable radiation patterns to structural parameters, bypassing laborious iterative simulations.[1] For light modulation and phase-gradient metasurfaces, DNNs model complex optical interactions between meta-atoms, enabling precise wavefront control for applications like beam shaping and holography, while significantly improving device efficiency compared to traditional phase-approximation methods.[2, 3] Scattering cross-section optimization and cloaking systems leverage DNNs to design metamaterials that suppress undesired scattering or achieve invisibility by dynamically tuning material properties under physical constraints.[4, 5] Near-field optical predictions, essential for plasmonic devices and cloaking platforms, are expedited through DNNs trained on simulation data, allowing rapid estimation of field distributions around nanostructures.[6, 7] Optical neural networks, inspired by biological systems, utilize nanophotonic components to mimic neural architectures, enabling energy-efficient, high-speed neuromorphic computing.[8] Reconfigurable metasurfaces further exemplify AI's adaptability, employing reinforcement learning to dynamically adjust meta-atom geometries or material states for real-time control of light polarization, phase, or amplitude.[9, 10]

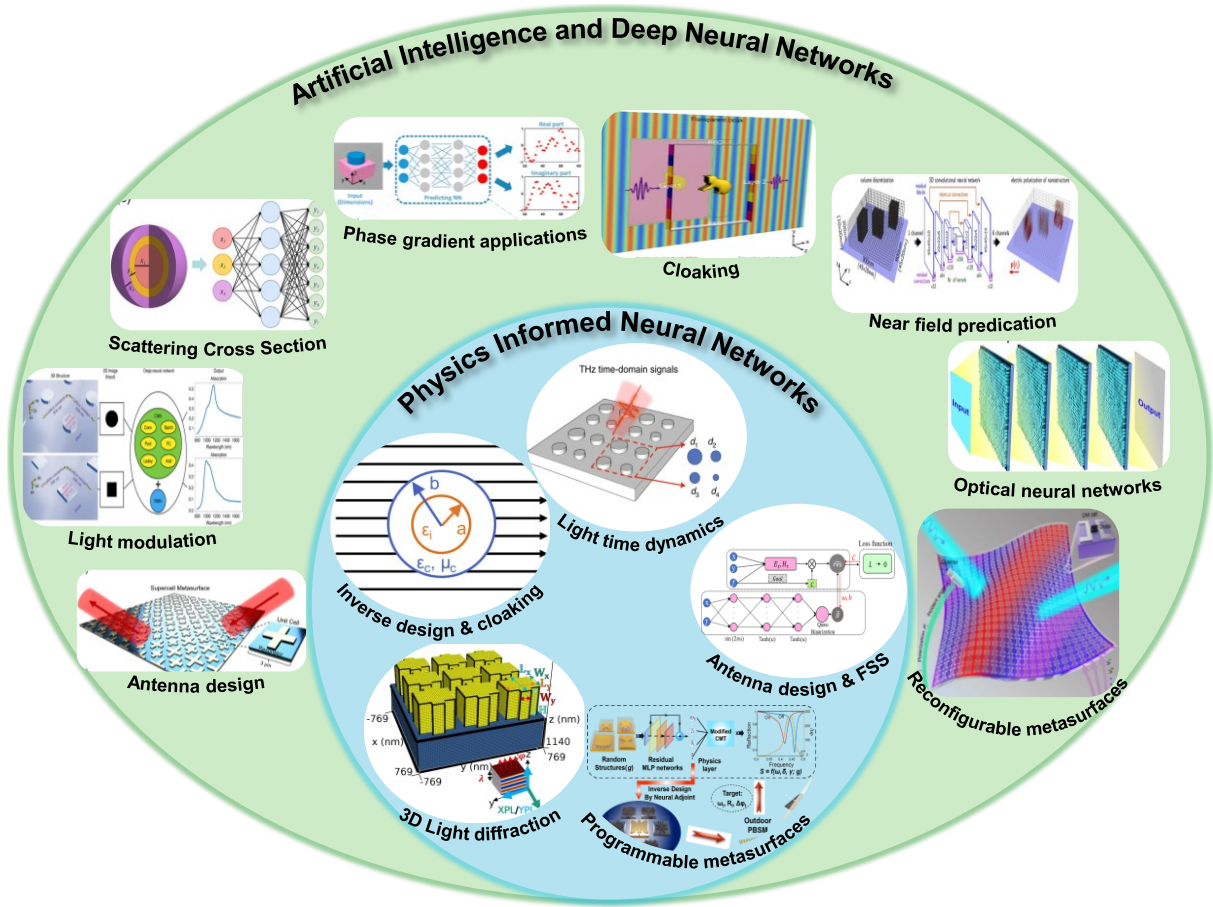

**Figure 1.** Integration of artificial intelligence, deep neural networks, and physics informed neural networks in nanophotonics and electromagnetism. (a) Applications of AI and DNNs in nanophotonics and electromagnetism: inverse design of frequency-selective surfaces (FSS) and antennas by linking electromagnetic responses to geometric parameters.[1] Reprinted with permission from De Gruyter, Copyright 2021. (b) light modulation via non-local coupling modelling in metasurfaces.[2] Reprinted with permission from Springer Nature, Copyright 2019. (c) scattering cross-section optimization for light scattering engineering.[4] Reprinted with permission from AAAS, Copyright 2018. (d) phase-gradient control for beam steering and holography.[3] Reprinted with permission from American Chemical Society, Copyright 2019. (e) cloaking and light scattering suppression.[7] Reprinted with permission from Optica, Copyright 2021. (f) near-field prediction around nanostructures.[6] Reprinted with permission from American Chemical Society, Copyright 2020. (g) optical neural networks for ultrafast neuromorphic computing.[8] Reprinted with permission from American Chemical Society, Copyright 2022. (h) reconfigurable metasurfaces with adaptive functionalities enabled by reinforcement learning.[9] Reprinted with permission from Springer Nature, Copyright 2020. Applications of physics-informed neural networks (PINNs). (i) inverse design and cloaking by embedding Maxwell's equations to optimize geometry and material distributions.[11] Reprinted with permission from Optica, Copyright 2020. (j) modelling light-time dynamics of terahertz pulses for real-time wave manipulation.[12] Reprinted with permission from Springer Nature, Copyright 2022. (k) antenna and FSS design through hybrid gradient-physics frameworks.[13] Reprinted with permission from ArXiv, Copyright 2024. (l) programmable metasurfaces with PINN-guided dynamic tuning of electromagnetic responses using phase-change or tunable dielectric materials.[14] Reprinted with permission from ArXiv, Copyright 2024. (m) 3D light diffraction analysis for holographic metasurfaces by solving wave-propagation equations.[15] Reprinted with permission from Optica, Copyright 2025.

Physics-informed neural networks (PINNs) unify data-driven learning with fundamental physical principles, offering a robust framework for solving complex challenges in nanophotonics and electromagnetism. By embedding physical governing equations such as Maxwell's equations or energy conservation laws into their architecture, PINNs reduce reliance on large training datasets while ensuring solutions adhere to physical constraints. In inverse design and cloaking, PINNs optimize non-intuitive geometries by minimizing scattering signatures or tailoring electromagnetic responses, enabling broadband invisibility and multifunctional device design.[11] For terahertz light dynamics, PINNs model ultrafast interactions between light and metasurfaces, advancing applications in next-generation communications and spectroscopy.[12] Three-dimensional light diffraction analysis benefits from PINNs' ability to solve wave-propagation equations, facilitating the design of holographic metasurfaces and complex optical systems.[15] Programmable metasurfaces, enhanced by PINNs, dynamically adapt their electromagnetic behaviour using tunable materials, achieving real-time reconfigurability for applications in adaptive optics and sensing.[14] In antenna and FSS design, PINNs combine gradient-based optimization with neural networks to generate high-performance devices with tailored radiation patterns, bridging computational efficiency with physical accuracy.[13] By integrating domain knowledge into AI frameworks, PINNs unlock novel paradigms for designing, controlling, and optimizing nanophotonic and electromagnetic systems, fostering innovation across both fundamental research and practical applications.

## 1.2. DNN fundamentals

DNNs represent a class of machine learning models characterized by multiple layers of nonlinear processing units. They have revolutionized the field of artificial intelligence by enabling systems to learn hierarchical representations directly from data.[16] This section introduces DNNs, discussing their architecture, learning mechanisms, and practical considerations.

DNNs are essentially extensions of traditional neural networks with an emphasis on depth, i.e., the number of hidden layers between input and output layers. Historically, neural networks began with few perceptron but limitations in expressivity and computational power confined early models to shallow architectures. The advent of efficient training algorithms (notably backpropagation) and advances in hardware have led to the resurgence and subsequent dominance of deep architectures. Today, DNNs underpin many breakthroughs in computer vision, speech recognition, natural language processing, and other domains.

### 1.2.1. Architectural Fundamentals

At the core of DNNs is the artificial neuron, a computational unit that mimics the behaviour of biological neurons. Each neuron computes a weighted sum of its inputs, adds a bias, and applies a nonlinear activation function:

$$a^{(l)} = f(W^{(l)}a^{(l-1)} + b^{(l)}) \quad (1)$$

where: $a^{(l-1)}$ is the vector of activations from the previous layer, $W(l)$ represents the weight matrix for layer $l$, $b^{(l)}$ is the bias vector, and $f(\cdot)$ denotes a nonlinear activation function (e.g., ReLU, sigmoid, or tanh).[17] The network is typically organized into an input layer (for receiving data), multiple hidden layers (for processing and feature extraction), and an output layer (for final prediction or classification). The term 'deep' in DNNs refers to the presence of several hidden layers. This hierarchical architecture enables the network to progressively construct abstract feature representations through layered learning mechanisms. First, initial few layers extract rudimentary local features (e.g., edges, textures). Second, intermediate layers synthesize these primitives into complex spatial or spectral patterns. Third, final layers encode high-level semantic descriptors tailored to the target application. Such tiered frameworks mirror the multiphysics design principles of functional metasurfaces, where subwavelength meta-atoms collectively enable macroscale wavefront engineering. This hierarchical abstraction proves indispensable for computationally intensive tasks, such as inverse design of dispersive optical antennas or multi-wavelength holography.

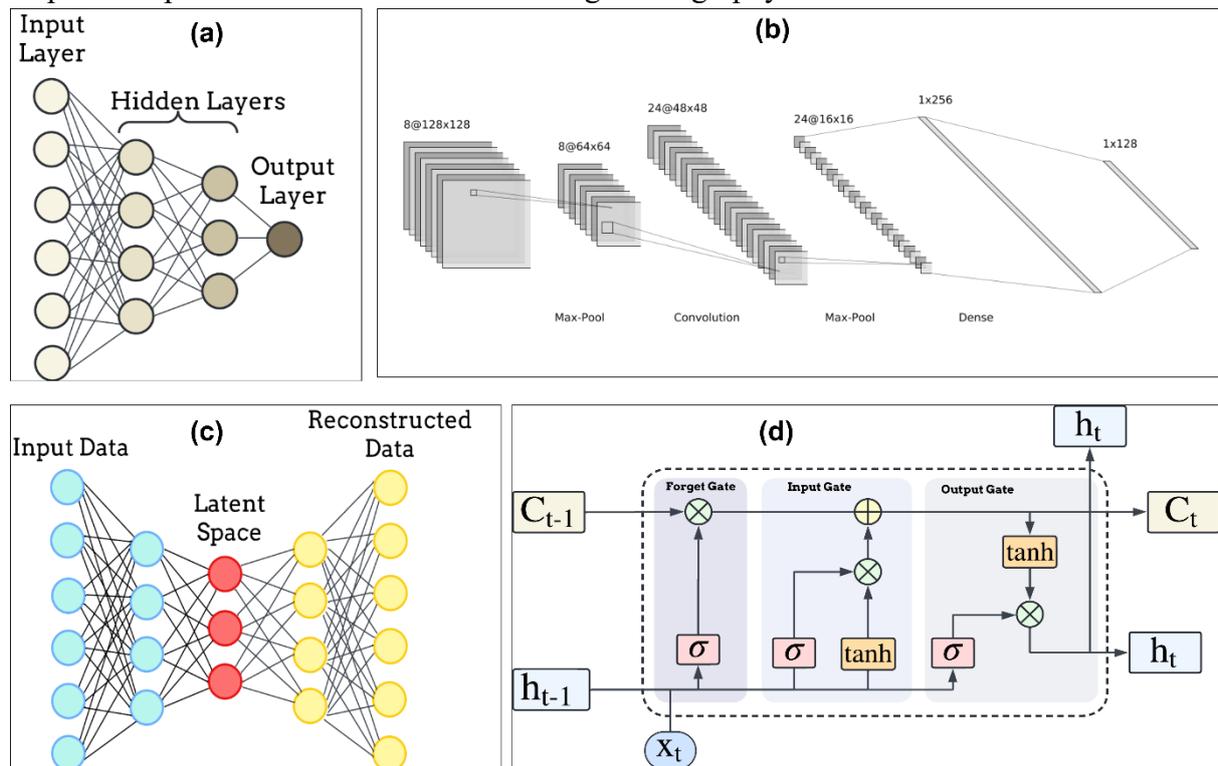

**Figure 2.** Overview of fundamental deep learning architectures used in various machine learning applications. (a) Standard DNNs network where information propagates through fully connected layers. (b) Convolutional neural networks (CNNs) are designed for spatial data processing, using convolutional layers to extract hierarchical features. (c) Autoencoder (AE) architecture, which compresses input data into a latent representation and then reconstructs it. (d) Long short-term memory (LSTM) network, incorporating gating mechanisms to capture long-range dependencies in sequential data.

### 1.2.2. Learning Mechanisms

**Forward Propagation:** During training, data is passed through the network in a process called forward propagation. Each layer transforms the input data using its weights and biases, generating activations that are passed to the subsequent layer. The final output is compared against the ground truth using a loss function, which quantifies the error in prediction.

**Backpropagation and Optimization:** Backpropagation is the cornerstone algorithm for training DNNs. It computes the gradient of the loss function with respect to each weight by propagating the error backward through the network. Mathematically, if $L$ denotes the loss, then the gradient for a weight $w$ is given by:

$$\frac{\partial L}{\partial w} = \frac{\partial L}{\partial a} \cdot \frac{\partial a}{\partial z} \cdot \frac{\partial z}{\partial w'} \tag{2}$$

where $z$ represents the pre-activation value. These gradients are then used to update the weights via an optimization algorithm (such as stochastic gradient descent, Adam, or RMSprop),[18] which iteratively minimizes the loss function.

**Regularization and Generalization:** To avoid overfitting, where the network memorizes the training data rather than learning generalizable patterns, various regularization techniques can be employed. Common methods include: (i) Dropout: Randomly deactivating a fraction of neurons during training to prevent co-adaptation; (ii) Weight Regularization: Adding L1 or L2 penalties to the loss function to constrain the magnitude of the weights; (iii) Early Stopping: Halting training once performance on a validation set starts to degrade.[19]

### 1.2.3. Variants and Extensions

While the basic feedforward deep neural network (DNN) serves as the foundation for various machine learning applications, specialized architectures have been developed to address specific data structures and learning tasks. Figure 2 illustrates four widely used deep learning models: Feedforward Neural Networks (FNNs), Convolutional Neural Networks (CNNs), Recurrent Neural Networks (RNNs) with Long Short-Term Memory (LSTM) units, and Autoencoders.

**Feedforward Neural Networks:** The simplest form is the deep neural networks (DNNs), consisting of fully connected layers where information propagates in one direction, from the input layer, through hidden layers, to the output layer (see Figure 2a). These networks are effective for structured data and function approximation problems but struggle with complex spatial or sequential dependencies. Despite their limitations, DNNs are foundational components in other advanced architectures.

**Convolutional Neural Networks (CNNs):** CNNs, depicted in Figure 2b, are specifically designed for handling spatial data such as images. They utilize convolutional layers that apply learnable filters to extract hierarchical features from input images. Pooling layers further down sample feature maps to reduce computational complexity while preserving essential patterns. CNNs have been instrumental in tasks such as image recognition, medical imaging, and object detection due to their ability to automatically learn spatial hierarchies.[20]

**Recurrent Neural Networks (RNNs) and LSTM Networks:** For sequential data, such as time series and natural language, RNNs introduce connections that allow information to persist across time steps. However, standard RNNs suffer from vanishing gradient problems, limiting their ability to capture long-term dependencies. LSTM networks (Figure 2c) address this issue by incorporating memory cells with gating mechanisms—forget, input, and output gates—to regulate information flow.[21] This enables LSTMs to learn long-range dependencies, making them suitable for speech recognition, text generation, and financial forecasting.

**Autoencoders and Generative Models:** Autoencoders, shown in Figure 2d, are a class of unsupervised learning models designed for dimensionality reduction, feature learning, and data generation. They consist of an encoder that compresses input data into a lower dimensional representation and a decoder that reconstructs the original input. Variants such as variational autoencoders (VAEs) and generative adversarial networks (GANs) extend this concept to generate realistic synthetic data.[22] These architectures play a crucial role in anomaly detection, image denoising, and representation learning. Each of these architectures builds upon the fundamental principles of deep learning while introducing specialized mechanisms to enhance performance for particular tasks.

## 1.3. Physics informed neural networks fundamentals

PINNs integrate DNNs mechanism with fundamental physical laws by embedding governing equations into the training process.[23] This combined approach enforce the learned solutions to be consistent with both underlying physical principles, and reduce the need for large datasets.[24,25] By unifying physics-based constraints with neural network approximations, PINNs provide a framework for solving complex linear and nonlinear problems where traditional data-driven methods may struggle to converge or predict the output properly.

### 1.3.1. Conceptual Framework

The primary idea behind PINNs is to leverage prior knowledge of physical systems to guide the learning process. Traditional deep neural networks learn from data alone, often requiring large datasets to capture the underlying dynamics accurately. In contrast, PINNs use the known structure of the physical laws to regularize the training.[26] This helps to: (i) reduce data dependency, where the embedded physics acts as an inductive bias, allowing the network to generalize well even with limited data;[27] (ii) enforce physical consistency by penalizing deviations from governing equations. In this way, PINNs yield solutions that satisfy known conservation laws and boundary conditions;[28] and (iii) facilitate inverse problems existing in design and optimization tasks, such as inverse design in nanophotonics, where PINNs can infer material properties or geometries that meet desired performance criteria by ensuring that the solutions obey physical constraints.[29]

### 1.3.2. Mathematical Formulation and Network Architecture

At the heart of a PINNs is the standard deep neural network, parameterized by weights $\theta$, which approximates a function $u(x, t)$ that represents the physical field of interest (e.g., the electromagnetic field). The key innovation is in how the training objective is constructed.

**Loss Function Augmentation:** A typical PINN loss function combines two primary components: First, data Loss ($\mathcal{L}_{data}$), this term measures the discrepancy between the network predictions and available observational or experimental data. Second, physics Loss ($\mathcal{L}_{PDE}$), while that term enforces that the network's predictions satisfy the governing partial differential equations (PDEs) of the physical system. Mathematically, the total loss $\mathcal{L}$ can be expressed as:

$$\mathcal{L} = \mathcal{L}_{data} + \lambda \mathcal{L}_{PDE} \qquad (3)$$

The parameter ($\lambda$) is the weighting parameter that balances the influence of the abstract data loss and physics loss. The partial differential equation of electromagnetic field is given by $\mathcal{N}[u(x,t)] = 0$, where $\mathcal{N}$ is a nonlinear differential operator. The physics loss is typically defined at a set of collocation points $[(x_i, t_i)]$ as:

$$\mathcal{L}_{PDE} = \frac{1}{N_c}\sum_{i=1}^{N_c} ||\mathcal{N}[u_\theta(\pmb{x_i},t_i)]||^2 \qquad (4)$$

This formulation forces the network's output $u_\theta$ to approximately satisfy the PDE at sampled points in the domain.

**Automatic Differentiation (AD):** step that computes the derivatives of physics quantities over space or time for the network output with respect to its inputs. These derivatives are essential for evaluating the differential operator $\mathcal{N}$ in the physics loss. PINNs avoids the need for numerical differentiation using AD approach.

**Training Procedure and Implementation Details:** Training process of PINN involves simultaneously reducing the data loss and physics loss to ensure accurate predictions compared with governing equations. The training process begins with data preparation, where the ground truth data are collected, and additional collocation points are chosen within the domain to enforce the physics constraints. Next, using the feedforward approach, the input data is propagated through the network to generate predictions $u_\theta(x,t)$. These predictions are then used in the error calculation step, where automatic differentiation is applied to evaluate the residuals of the governing equations, ensuring compliance with the prescribed physical laws. Then, the total loss including both data-driven and physics-informed loss feeds into a single objective function, balancing the influence of data and constraints. Finally, backpropagation and optimization process are performed using gradient-based methods such as Adam or other proper method to iteratively refine the network weighting parameters $\theta$. This structured framework ensure the network will learn solutions that adhere to both empirical observations and fundamental physical principles using fewer input data, which makes PINNs a powerful tool for physics-constrained learning.

### 1.3.3. Emerging PINN architectures

The integration of PINNs is particularly useful in complex scientific and engineering problems where data is limited, and traditional numerical methods are computationally expensive especially for big systems. Figure 3a demonstrates a PINN framework designed to solve Maxwell's equations, where separate neural networks model the electromagnetic field and material parameters, enforcing physical constraints directly within the learning process.[30] Convolutional neural networks alongside finite difference methods has been used to efficiently capture spatial dependencies while minimizing PDE residuals in Figure 3b.[31] Physics-informed autoencoders can extract meaningful latent representations while preserving physical consistency, making them valuable for dynamical system modelling Figure 3c.[32] The use of an LSTM-based PINN for solving time-dependent differential equations, ensuring that learned temporal dynamics remain physically plausible Figure 3d. These recent approaches showcase the versatility of PINNs in integrating physics with neural networks, enhancing accuracy, generalization, and interpretability across diverse applications.[33]

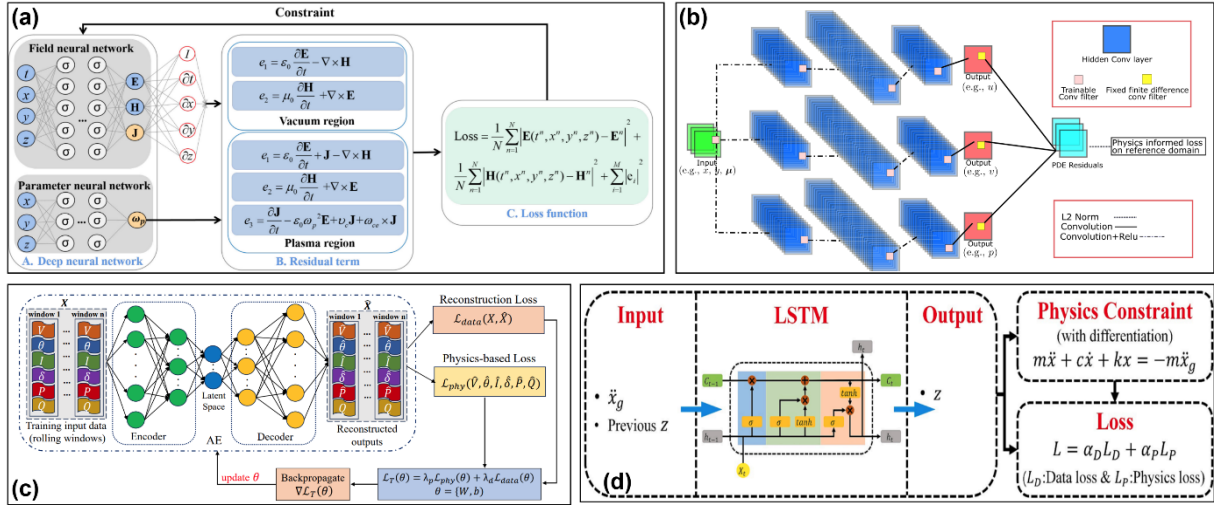

**Figure 3.** Recent PINN architectures with physical constraints into DNN learning. (a) PINN with DNN solves Maxwell's equations with separate field and parameter networks. Reprinted with permission from ArXiv, Copyright 2020. (b) A convolutional PINN incorporating finite difference methods to enforce PDE residuals. Reprinted with permission from ArXiv, Copyright 2024. (c) A physics-informed autoencoder (AE) leveraging latent-space representations and physics-based loss. Reprinted with permission from Elsevier, Copyright 2020. (d) A physics-informed LSTM capturing time-dependent dynamics with embedded differential equation constraints. Reprinted with permission from Elsevier, Copyright 2023.

# 2. Artificial Intelligence for Nanophotonics Design

## 2.1. Nanophotonics Overview

### 2.1.1. Fundamental principles

Nanophotonics represents a transformative frontier in the manipulation and understanding of light-matter interactions at nanometers scales, bridging the classical boundaries of bulk optics with the innovative potentials of nanoscale materials and structures.[34] The field capitalizes on the unique behaviours of light when confined to dimensions comparable to its wavelength, offering unprecedented control over optical properties and functionalities. Unlike conventional bulk optics, which relies on natural materials with fixed refractive indices and straightforward phase manipulation, nanophotonics leverages engineered materials, called metamaterials and metasurfaces, to achieve tailored electromagnetic responses that are unattainable in nature.[35]

Bulk optics has historically been rooted in the manipulation of light through macroscopic lenses, prisms, and mirrors, governed by well-established principles such as Snell's law and Fresnel equations. These components control light primarily through gradual phase accumulation over propagation distances significantly larger than the wavelength of light. In contrast, metamaterials explore the interaction of light with subwavelength structures, where phase, amplitude, and polarization can be controlled at the nanoscale through abrupt phase discontinuities. Including the phase of light in reflection law results in generalized Snell's laws. New laws extend traditional refraction and reflection equations by incorporating engineered phase gradients as in equations 1 and 2.[36] Metasurfaces, two-dimensional version of bulk metamaterials enable an emerging field of flat optics for future compact nano lenses "metalenses" and compact LIDAR and nano-light sources as plotted in Figure 4a.[37]

$$\sin(\theta_r) - \sin(\theta_i) = \frac{\lambda_o}{2\pi n_i}\frac{d\phi}{dx} \qquad (5)$$

$$\sin(\theta_t)\,n_t - \sin(\theta_i)\,n_I = \frac{\lambda_o}{2\pi}\frac{d\phi}{dx} \qquad (6)$$

The fundamental physics of metasurface revolves around light scattering, resonant phenomena, and subwavelength field confinement. Mie theory explains theoretically the behaviour of light scattering with designed metasurface at specific electromagnetic wavelength. Other phenomena called localized surface plasmon resonances and describes the light confinement in metallic nanostructures. The Fano resonance, characterized by asymmetric line shapes arising from interference between a discrete state and a continuum, further highlights the rich interplay of light with engineered nanostructures.[38] Such resonances enable strong optical absorption for solar cells.[39-52]

One of the groundbreaking concepts in nanophotonics is phase gradient light refraction, where metasurfaces impart spatially varying phase shifts to manipulate the trajectory of light. This principle enables the design of ultrathin optical elements, such as flat lenses and beam deflectors, which outperform their bulk counterparts in terms of compactness and integration potential. By engineering the phase profile across a metasurface, light can be bent or focused with unprecedented precision, opening pathways for applications in nonlinear optics, structural colors, data communication, and quantum optics.[53-60]

In this section, we will review Nanophotonics fundamentals and popular design methods of nanophotonics. The application of AI in nanophotonics design for forward and inverse design. Lastly, the integration of physical optics to refine the design capability of AI for applications in nonlinear optics, accelerated design, and high-dimensional optimization such as optical neural networks.

### 2.1.2. Traditional design methods

In nanophotonics, precise simulation of light-matter interactions at the nanoscale is essential for designing and optimizing devices. Traditional computational methods like Finite-Difference Time-Domain (FDTD) and Finite Element Method (FEM) are widely employed for this purpose, each offering distinct advantages and limitations.[61, 62]

FDTD is a time-domain method that solves Maxwell's equations by discretizing both time and space. It computes the electromagnetic field evolution over time, making it particularly effective for modelling broadband responses and transient phenomena. FDTD's straightforward implementation and ability to handle complex, inhomogeneous materials contribute to its popularity in nanophotonics. However, achieving high accuracy necessitates fine spatial and temporal discretization, which can lead to substantial computational demands, especially for large-scale simulations.[63]

FEM operates in the frequency domain by discretizing the computational domain into finite subdomains, where electromagnetic fields are approximated using basis functions. This approach enables precise modelling of complex geometries and material heterogeneities, making it particularly advantageous for capturing localized field distributions in nanophotonic systems. However, FEM's reliance on solving large systems of equations can lead to significant computational overhead, especially for problems involving extensive domains or broadband frequency analyses.

In nanophotonics, the selection of computational methodologies critically influences simulation accuracy and efficiency. FDTD method employs explicit time-stepping to resolve transient electromagnetic interactions. FDTD offers efficient parallelization for time-domain applications. Despite its computational speed, FDTD often requires fine spatial discretization near material interfaces to mitigate numerical artifacts. FDTD has a huge memory demands especially for large-scale three-dimensional simulations.[64]

On the other hand, FEM utilizes an implicit formulation, which leads to longer simulation times for large-scale problems. Nonetheless, FEM's ability to conform to complex geometries often provides superior accuracy in modelling intricate structures. Its memory requirements are closely tied to the complexity of the mesh and the order of the basis functions used, potentially offering more efficient memory utilization for certain problems.

Regarding application suitability, FDTD is well-suited for simulating broadband phenomena, such as pulse propagation and time-resolved studies, due to its time-domain nature. It is also effective for modelling periodic structures and photonic crystals. In contrast, FEM is ideal for frequency-domain analyses, particularly when dealing with complex geometries, anisotropic materials, or scenarios requiring high precision in localized field distributions.[65]

## 2.2. Design Nanophotonics using DNN

### 2.2.1. Forward design of nanophotonics devices

The forward design of nanophotonic devices involves predicting the optical responses, such as light scattering, transmission, reflection, or absorption spectra, based on given geometrical and material parameters. DNNs have emerged as a transformative tool for this purpose, providing a data-driven alternative to traditional computational methods like finite-difference time-domain (FDTD) or finite element methods (FEM). Unlike these numerical approaches, DNNs can significantly reduce computation times while maintaining high accuracy, thus accelerating the iterative design and discovery of novel nanophotonic devices.[66-69] This section reviews recent advancements in applying DNNs to the forward design problem, highlighting key methodologies and innovations.

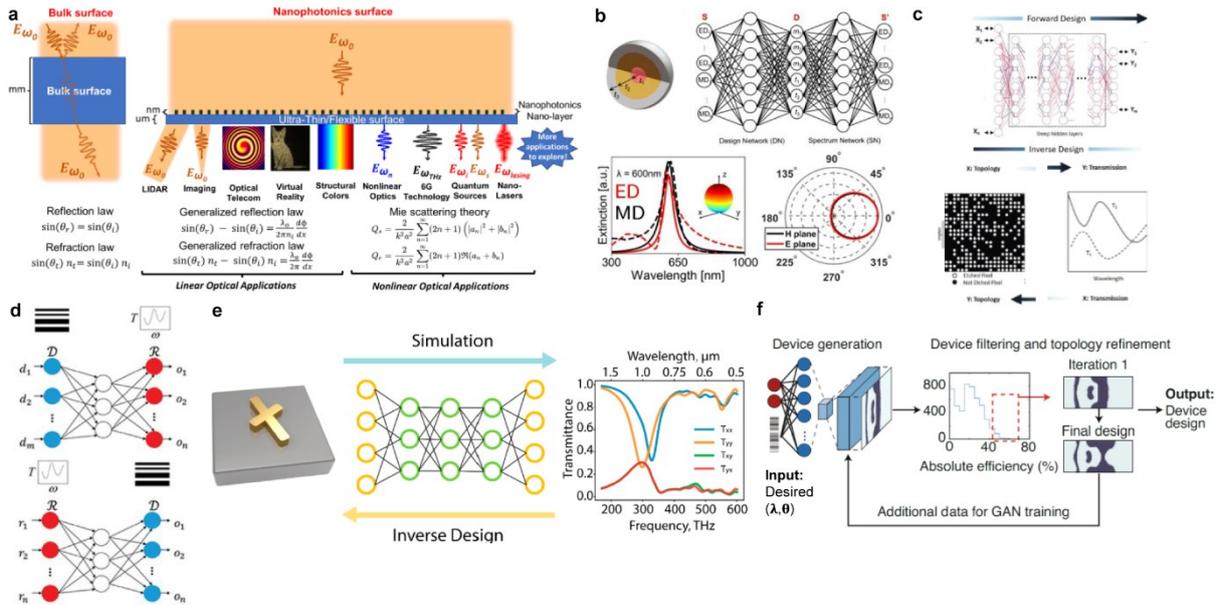

**Figure 4.** Applying DNNs in Nanophotonics for forward and inverse design. (a) Comparison between bulk optics and nanophotonics light-matter interactions. (b) Designing a desired light scattering in the far field from multi-layer nano-scatter using DNNs.[4, 70] Copyright 2019 American Chemical Society. (c) Optimizing design of integrated photonic power splitter using DNNs.[71] Copyright 2019 Springer Nature. (d) Light modulation using DNN for forward and inverse design of a multilayer $SiO_2$-$Si_3N_4$ cavity.[72] Copyright 2018 American Chemical Society. (e) Accelerating plasmonic metasurface design using generative model for forward and inverse approach.[73] Copyright 2018 American Chemical Society. (f) Inversely getting a high diffraction efficiency at certain wavelength from nanophotonics devices using generative adversarial networks.[74] Copyright 2019 American Chemical Society.

One of the most notable early applications of DNNs in nanophotonics forward design was the use of fully connected neural networks to predict the spectral responses of plasmonic and dielectric nanostructures. Malkiel *et. al.* demonstrated a neural network architecture trained on simulated datasets to predict the far-field scattering spectra of plasmonic nanoparticles. The network, once trained, achieved prediction times orders of magnitude faster than traditional solvers. This work established the feasibility of applying DNNs to forward design tasks and inspired subsequent research to enhance accuracy and versatility.[75]

Building on this, Peurifoy *et. al.* used multilayer DNNs to predict the optical spectra of multilayered nanostructures. Their work introduced a robust training protocol using synthetic datasets generated via FDTD simulations, ensuring the model could generalize across a wide range of design geometries for getting on demand light scattering pattern in the far field. The simplicity and scalability of their approach made it applicable to a broad class of nanophotonic devices.[4, 70] Mohammed *et. al.* developed DNNs that trained on extensive simulation data to accurately predict the spectral response of complex nanophotonic structures and to design optimized components such as compact silicon-on-insulator (SOI) power splitters with targeted splitting ratios. The model demonstrated high efficiency, achieving over 90% transmission, minimal reflection, and precise power splitting specifications within seconds. This approach highlights the potential of DNNs to handle large design spaces, significantly accelerating the development of advanced nanophotonic devices and paving the way for their broader application in integrated photonics.[71]

An improved deep learning method for inverse design was developed by Liu *et. al.* through utilizing a tandem DNNs architecture that integrates both forward model with an inverse design network. This approach was applied to design $SiO_2$ and $Si_3N_4$ multilayer structures, with layer thicknesses as the design parameters. The forward model, trained on 500,000 labelled data pairs, maps design parameters to transmission spectra. The backpropagation training minimizes the loss between the desired and predicted spectra. The developed architecture addresses non-uniqueness by focusing on minimizing spectral differences rather than constraining specific design parameters. The method was further extended to design 2D structures capable of modulating transmission phase delay at multiple wavelengths, demonstrating its versatility in nanophotonic device design.[72]

As device geometries became huge and complex, CNNs were explored for their ability to capture spatial relationships within nanostructures. Deep learning-based approach has improved $Q$-factor of two-dimensional photonic crystal (2D-PC) nanocavities through dimensional optimization. Asano *et. al.* used a CNNs trained on a dataset of randomly generated nanocavities, the method efficiently established a relationship between structural variations and $Q$-factors. This enabled rapid gradient-based optimization and resulted in a nanocavity design with a $Q$-factor dramatically higher than manual optimization methods. That approach demonstrates the potential of combining deep learning with structural optimization to achieve superior optical performance.[76] On the other hand, So *et. al.* applied CNNs to predict the far-field response of metasurfaces with non-periodic arrangements. Their work utilized 2D geometric representations of the structures as input, enabling the model to learn spatial correlations and improve accuracy for irregular geometries.[77] Another work reported by Wiecha *et. al.*, combined CNNs with a feature extraction pipeline of complex 3D nanostructures. The developed model was trained to predict multiple spectral responses, such as transmittance and reflectance. The novelty of this work lay in its generalization technique which able to handle diverse geometries without requiring specialized preprocessing.[6]

Encoder-decoder architectures have strong advantage in the forward design of nanophotonic devices, which requires high-dimensional input-output mappings. Tan *et. al.* proposed a U-Net-inspired architecture to predict the spectral response of metasurfaces directly from pixelated images of their geometric layouts. By using a hierarchical encoding scheme, the network captured fine-grained geometric details and translated them into accurate spectral predictions.[78] This approach was further extended by Gao *et. al.*, who integrated attention mechanisms into the encoder-decoder model. The attention modules allowed the network to focus on critical regions of the device geometry to improve the prediction accuracy for highly anisotropic or heterogeneous structures. The study demonstrated the efficacy of combining spatial feature extraction with interpretability and provide insights into the physical mechanisms driving the predicted optical responses.[79]

One of the challenges in applying DNNs to forward design is the need for large and high-quality training datasets. To address these limitations, Ma *et. al.* introduced a semi-supervised learning framework that leveraged a small number of labelled samples alongside unlabelled data to enable data-efficient learning strategies. Their model incorporated self-supervised pretraining to learn geometric features before fine-tuning on specific spectral prediction tasks. This approach reduced the reliance on extensive simulation datasets while maintaining high prediction accuracy.[80]

Transfer learning has also been applied to mitigate data scarcity. Chen *et. al.* (2022) trained a neural network on a large dataset of generic nanostructures and fine-tuned it for specific device geometries. This methodology not only shortened training times but also improved the generalization capabilities of the model, making it adaptable to diverse design requirements.[81]

More recently, Adibnia *et. al.* (2023) developed a recurrent neural network (RNN)-based model to handle spectral data as a sequence, enabling it to capture correlations across adjacent wavelengths. By treating the spectral response as a time-series problem, the model achieved superior accuracy compared to traditional fully connected networks. This innovation opened new possibilities for designing devices with complex dispersion properties.[82]

One of the ultimate goals of forward design using DNNs is real-time prediction to facilitate interactive design workflows. Gao *et. al.* (2022) achieved this by deploying lightweight neural network models optimized for inference speed. Their approach involved pruning redundant layers and parameters from a pre-trained network to reduce its size without compromising accuracy. The resulting model was capable of predicting optical responses in milliseconds and enable real-time feedback for iterative design.[83]

By utilizing unsupervised learning, the model design by Liu *et. al.* autonomously discovers and optimizes metasurface structures, eliminating reliance on human expertise and iterative parameter sweeping. This approach efficiently handles complex design tasks, such as multiple input spectra or gradient structural distributions, and accelerates the discovery of novel optical phenomena. Future enhancements, including advanced network configurations and physically meaningful loss functions, could further improve its performance and versatility in photonic device design.[73]

Jiang *et. al.* introduced an innovative inverse design method by combining generative adversarial networks (GANs) with topology optimization to design metasurfaces with high diffraction efficiency at target wavelengths and incident angles. This approach leverages GANs to efficiently generate training data and capture essential features of metasurface designs, while topology optimization further enhances device performance. By focusing on extracting critical design features rather than predicting optical properties. The developed method significantly reduces the need for extensive training datasets and offers a computationally efficient strategy for creating high-performance metasurfaces.[74]

The forward design of nanophotonic devices has been improved using DNNs through enabling fast and accurate prediction of optical responses, reducing computational costs, and expanding design possibilities. Using different architectures such as convolutional architectures, encoder-decoder frameworks, generative adversarial networks, and data-efficient strategies have enhanced design efficiency and precision. However, challenges remain in improving generalization to unseen data, reducing data requirements, and ensuring robustness under physical constraints. Future directions include integrating domain knowledge, such as embedding Maxwell's equations into training process. Moreover, developing hybrid models that combine neural networks with traditional solvers will enhance accuracy and versatility and paves the way for designing next-generation nanophotonic devices.

### 2.2.2. Inverse design of nanophotonic structures

Combining DNNs with the inverse design of nanophotonic devices has significantly advanced the field and enabled the development of complex structures with tailored optical properties. This section reviews notable contributions in this area, and highlights the unique methodologies and innovations introduced by various research efforts.

Tahersima *et. al.* deployed DNNs with integrated nanophotonics devices to predict the optical response of engineered nanophotonic geometries. Their model facilitated inverse design to achieve the targeted optical characteristics of compact silicon-on-insulator (SOI) 1×2 power splitters. The DNN successfully achieved various splitting ratios with high transmission efficiency (above 90%) and minimal reflection (below -20 dB). The improved design compared with other design methods demonstrated the potential of using DNNs in rapidly designing integrated photonic components with complex nanostructures.[71]

Using the same platform, Kojima *et. al.* explored three DNN models for designing nanophotonic power splitters with multiple splitting ratios. First, forward regression model predicts the spectral response given a specific structural design. Second, inverse regression model generates a structural design based on a desired spectral response. Third, generative model produces a series of optimized designs corresponding to target performance metrics. The study showcased how these DNN models could efficiently navigate the design space to produce devices meeting specific performance criteria.[84]

A method combining a generative model and a genetic algorithm was proposed by Zhu *et. al.* to overcome the problem of uncertainty in design. This approach compresses real spectra into a latent space using the generative model. Then, decodes target spectra based on performance indicators, and employs a hybrid optimization algorithm combining a genetic algorithm with a forward prediction network to refine the generated spectra. The proposed process enables the systematic inverse design of structural parameters from target spectra and demonstrated experimentally through the on-demand design of multilayer nanofilms. The method offers a promising solution for achieving inverse design of nanophotonic devices tailored to specific performance requirements.[85]

Integrating deep learning methods with an enhanced recurrent neural network (RNN) can capture the sequence characteristics of a spectrum for inverse design and prediction. The network's memory and feedback loops enable it to effectively process time-series data, demonstrating high consistency between target and predicted spectra in the context of nanorod hyperbolic metamaterials. The model propose by Yan *et. al.*, which also performs well with user-drawn spectra, can predict unknown spectra with only a 0.32% mean relative error, offering a promising tool for fast and accurate nanophotonic device design. This method provides a more efficient alternative to traditional ANN applications in nanophotonics, enabling the precise prediction of spectra beyond detection limits.[86]

Frising *et. al.* addressed the challenge of multimodal distributions in inverse photonic design, where multiple device configurations can yield similar performance. They implemented a Conditional Invertible Neural Network (cINN) to provide the full distribution of possible solutions, resolving ambiguities inherent in multimodal device distributions. Unlike conventional approaches, generative models provide a full distribution of possible solutions with multiple valid designs. This work compares two generative models, cVAE and cINN on a nanophotonic problem involving transmission spectrum tailoring through subwavelength indentations in a metallic film. Results show that cINNs offer superior flexibility over cVAEs, particularly when dealing with multimodal device distributions. cINNs is a promising tool for overcoming challenges in inverse design.[87]

The integration of DNNs into the inverse design of nanophotonic devices has led to significant advancements in device performance and design efficiency. Developed models predict and design optical responses to addressing challenges like multimodal distributions and data inconsistencies. As deep learning techniques continue to evolve, their applications in nanophotonics promises increasingly sophisticated and efficient design methodologies.[87-90]

### 2.3. PINNs for Nanophotonics

The integration of artificial intelligence (AI), particularly PINNs, offers a promising avenue to overcome challenges in accuracy and computational efficiency of pure DNN. PINNs incorporate physical laws directly into the neural network architecture, enabling efficient and accurate modelling of complex systems.[60] This section reviews recent advancements in applying PINNs to model the optical properties of nanophotonic devices, highlighting the novel approaches and contributions of each work.

## 2.3.1. Modelling optical properties of nanophotonics devices

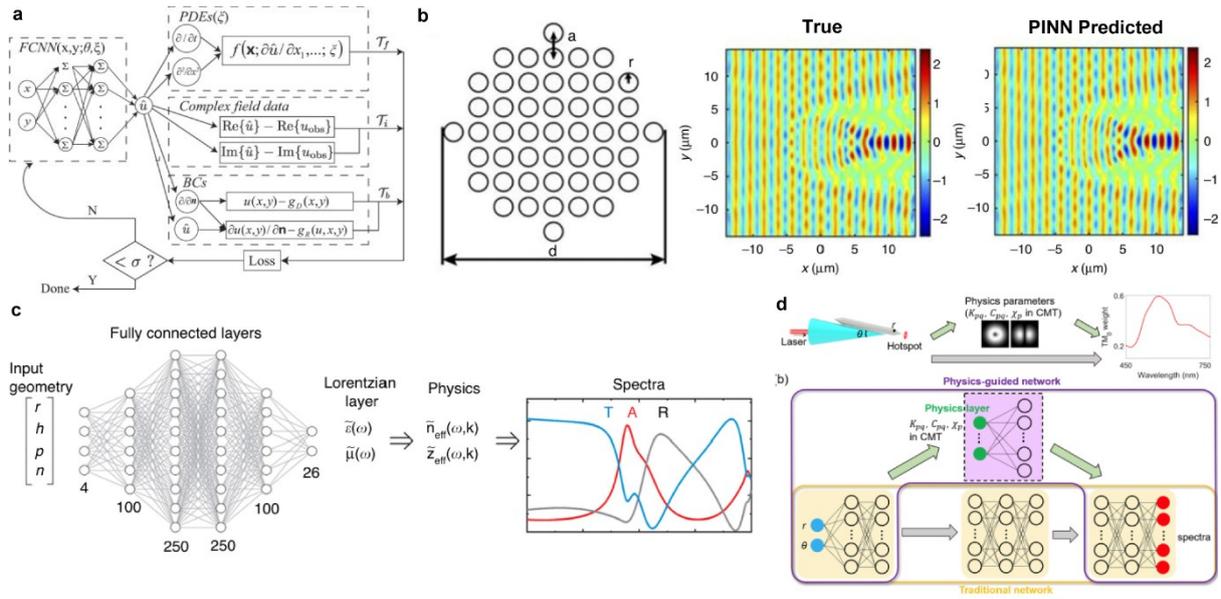

**Figure 5.** PINN applications in Nanophotonics for forward and inverse design. (a) Flow chart of PINN for inverse design and retraval of near field microscopy that consists of operators of partial differential equations, field observations, and boundary conditions.[91] Copyright 2022 AIP publishing LLC. (b) PINN model for optical properties retrieval from near field light scattering.[11] Copyright 2020 Optica publishing group. (c) Lorentz neural network for forward design of nanophotonics devices.[92] Copyright 2022 Wiley-VCH. (d) PINN for forward design of hybrid photonic-plasmonic integrated photonic waveguide.[93] Copyright 2022 American Chemical Society.

Chen *et. al.* introduced the application of PINNs to solve inverse scattering problems in photonic metamaterials and nano-optics. Their approach involved retrieving effective permittivity parameters of finite-size scattering systems, including multi-component nanoparticles. By embedding Maxwell's equations into the neural network's loss function, the PINNs could infer material properties from scattered field data without requiring extensive datasets. This method demonstrated the potential of PINNs to handle complex inverse problems in nanophotonics, offering a mesh-free alternative to traditional numerical methods.[11]

Building upon their previous work, Chen and Dal Negro developed a PINN framework for imaging and parameter retrieval of photonic nanostructures from near-field data. Their method utilized full-vector Maxwell's equations to inversely retrieve spatial distributions of complex electric permittivity and magnetic permeability. The study showcased the capability of PINNs to achieve accurate convergence to true material parameters under various excitations, including plane waves and localized sources. This advancement is particularly relevant for techniques like scanning near-field optical microscopy (SNOM), where precise material characterization is crucial.[91]

Wang *et. al.* introduced a multi-receptive-field physics-informed neural network designed to acquire the electromagnetic response of intricate media at the nanoscale. Their approach addressed the challenges in designing nanophotonic devices through modelling light interactions within the media. The novelty of their work lies in the integration of multiple receptive fields within the PINN architecture, enabling the capture of both local and global features of the electromagnetic fields. Their method enhances the accuracy of simulations and facilities the design of more efficient nanophotonic devices.[94]

Lorentz deep neural network (LNN) has been developed by Khatib *et. al.* to learn the underlying physics of all-dielectric metamaterials. By incorporating a physics-based Lorentz layer that enforces causality, the approach establishes a relationship between geometrical parameters and the frequency-dependent permittivity and permeability of the material. Additionally, the LNN determines spatial dispersion and provides insights into both complex and dispersive permittivity and permeability. Relaying on trained data from electromagnetic simulations, the LNN autonomously produces effective material parameters, and demonstrates strong alignment with theoretical and simulated results. This method is applicable beyond metamaterials, making it a significant step toward integrating neural networks with physical laws for a wide range of material systems.[92]

A physics-guided, two-stage machine learning network has been developed by Liang *et. al.* to address challenges in designing hybrid photonic-plasmonic devices for efficient nanoscale light-matter interaction. By leveraging improved coupled-mode theory for optical waveguides, the network enhances the predictive accuracy of optical device designs to 98.5% while overcoming the input-output dimension mismatch. This approach enables near-unity coupling efficiency with symmetry-breaking selectivity, as demonstrated by the fabrication of photonic-plasmonic couplers with predicted profiles. These couplers achieve an excitation efficiency of 83% for radially polarized surface plasmon modes, offering a pathway toward super-resolution optical imaging with reduced computational costs.[93]

### 2.3.2. Accelerating Nanophotonics Simulations

By embedding physical laws such as Maxwell's equations directly into the neural network architecture, PINNs reduce the dependency on large datasets typically required for training conventional models. This integration not only accelerates simulations but also ensures adherence to underlying physical principles, enhancing model reliability. PINNs have demonstrated remarkable efficiency in solving forward problems, where optical responses are predicted from structural parameters, and inverse problems, where desired optical properties guide the design of nanophotonic structures. Their ability to optimize computational resources and achieve rapid convergence makes PINNs a promising approach for tackling complex, high-dimensional problems in nanophotonics with unprecedented speed and accuracy.[95]

WaveY-Net, a hybrid data- and physics-augmented convolutional neural network developed by Chen *et. al.*, revolutionizes nanophotonics design by enabling ultrafast and accurate predictions of electromagnetic field distributions for dielectric nanostructures. This method trains on magnetic near-field distributions and incorporates Maxwell's equations to compute electric fields and enforce physical constraints, achieving high accuracy while dramatically reducing computational demands. WaveY-Net significantly accelerates both local and global optimization processes for diffractive photonic devices, reducing design cycle times by orders of magnitude compared to traditional methods. Its data-efficient training approach enhances generalization, making it particularly advantageous for large-scale three-dimensional systems where generating extensive training datasets is computationally prohibitive. By integrating seamlessly with gradient-based algorithms, WaveY-Net transforms the design landscape, enabling rapid exploration of complex parameter spaces and efficient prototyping of high-performance nanophotonic devices. While tailored for fixed topologies and materials, its adaptability suggests potential for broader applications, including ensemble approaches for diverse photonics problems, thereby paving the way for a new era of accelerated and scalable nanophotonics design.[96]

Another approach developed by Park *et. al.* called Physics-informed reinforcement learning (PIRL), which integrates the adjoint-based method with reinforcement learning (RL) to significantly accelerate nanophotonic design by improving sample efficiency by an order of magnitude compared to conventional RL and overcoming local minima challenges. This approach, demonstrated on one-dimensional metasurface beam deflectors, achieves superior performance, exceeding most reported records in the field. PIRL leverages transfer learning to further enhance efficiency by adapting networks across different design problems, and reward engineering enforces practical constraints, such as minimum feature size, ensuring fabrication feasibility. Additionally, PIRL's compatibility with simulation tools that compute local design gradients, such as automatic differentiation-enabled RCWA tools, extends its applicability to a wide range of complex devices, including meta-gratings and metalenses. By combining physical insights with advanced RL techniques, PIRL provides a robust and scalable framework for optimizing intricate nanophotonic devices, enabling faster, more efficient exploration of vast design spaces and addressing previously intractable challenges in freeform device design.[97]

Optimizing the nanophotonic design to include inhomogeneity and anisotropy of the scattering medium surrounding nano-scatters was challenging till Wang *et. al.* proposed the multi-receptive-field physics-informed neural network (MRF-PINN), which represents a significant advancement in accelerating nanophotonic design by efficiently solving electromagnetic field problems in complex nano-scattering media, including dispersion, inhomogeneity, anisotropy, nonlinearity, and chirality. This innovative framework employs six encoder-decoder modules with variable receptive fields and a weighted aggregation scheme to accurately capture electromagnetic perturbations. By integrating physical constraints through informed loss functions and incorporating a scale balancing algorithm, MRF-PINN ensures rapid network convergence while maintaining high precision. Empirical results demonstrate that MRF-PINN can reconstruct electromagnetic field distributions within tens of milliseconds, providing quasi real-time analysis and eliminating the need for computationally intensive modelling of diverse nanostructures. This work highlights the potential of MRF-PINN as a scalable and cost-effective solution for forward and inverse design scenarios in nanophotonics.[94]

Appartently, PINNs and DNNs have unique approach in predicting nanophotonic design. Since PINNs inherently incorporate physical laws to ensure that simulations adhere to fundamental principles. This enhances speed, accuracy and generalization even with limited data. On the other hand, DNNs lacking explicit physical constraints, and may require larger datasets for acceptable prediction efficiency and have a big risk producing unphysical results. Furthermore, PINNs have demonstrated robustness in handling noisy data and complex parameter retrieval tasks, which outperformed DNNs in various studies. Although DNNs offer computational speed-ups as surrogate models, the integration of physics in PINNs provides a more reliable and efficient approach for accelerating nanophotonic design simulations.[11]

### 2.3.3. Modelling Nonlinear Optical Phenomena

PINNs have been utilized in predicting performance of nonlinear optical systems through integrating the governing equations of physics into their neural network architectures. In the field of nanophotonics, nonlinear optical phenomena such as self-focusing, harmonic generation, and soliton dynamics are pivotal for advanced device functionalities. These physical phenomena are governed by complex nonlinear equations, including the nonlinear Schrödinger equation (NLSE), and other high-dimensional PDEs. Traditional numerical methods for solving such problems are computationally intensive and scale poorly with increasing complexity. PINNs address these challenges by providing a data-efficient, physics-constrained framework to approximate solutions of nonlinear equations efficiently. By leveraging the inherent physical laws of nonlinear optics, PINNs enable the accurate prediction of optical responses in intricate nanostructures, and reduce computational demands significantly. This approach not only accelerates the modelling and simulation process but also facilitates the exploration of novel nonlinear optical phenomena and device designs, such as soliton-based waveguides and harmonic-generating metasurfaces, thereby advancing the capabilities of nanophotonic technologies.[98-100]

Liu *et. al.* designed a PINNs that are accurately predicting the dynamics of self-trapped necklace solutions of the (2+1)-dimensional nonlinear Schrödinger/Gross-Pitaevskii equation. These necklace patterns, characterized by integer, half-integer, and fractional reduced orbital angular momenta, exhibit unique rotational behaviours akin to rigid body motion and centrifugal force. Unlike ordinary fast diffraction-dominated dynamics, these patterns maintain quasi-stable propagation over multiple diffraction lengths despite gradual expansion or contraction. PINNs effectively emulate the solution of nonlinear PDEs and capture intricate dynamical properties. Moreover, PINNs demonstrate robustness through parameter discovery under both clean and noise-perturbed data, which enable precise predictions of nonlinear dynamics. This approach underscores the potential of PINNs to revolutionize the modelling of nonlinear optical phenomena and complex light-matter interactions in nanophotonic devices.[101]

PINNs has been used to model nonlinear optical dynamics through solving the nonlinear Schrödinger equation for fiber optics applications. PINNs accurately simulate multiple physical effects in optical fibers, including dispersion, self-phase modulation (SPM), and higher-order nonlinear phenomena. By embedding physical parameters such as pulse peak power and sub-pulse amplitudes as input controllers, PINNs achieve robust generalizability across various scenarios for soliton and multi-pulse propagation. Compared to traditional split-step Fourier methods (SSFM) and DNNs, PINNs require fewer computational resources and significantly less data while maintaining high accuracy. This capability enables PINNs to bridge the gap between computational efficiency and physical accuracy and make PINNs highly suitable for modelling nonlinear optical phenomena in nanophotonics. Also, their potential extends to applications such as optical signal processing, photonic material design, and parameter identification.[98]

Another significant application is in predicting nonlinear optical scattering. Traditional methods for modelling scattering in nonlinear media often involve solving complex equations that can be computationally demanding. A work introduces MaxwellNet, a physics-driven DNN designed model. It predicts nonlinear optical scattering in the presence of the optical Kerr effect, specifically for microscopic objects with sizes comparable to the incident wavelength. By incorporating a tunable network with respect to incident power, MaxwellNet can model intensity-dependent refractive index variations and perform topology optimization of nonlinear optical devices, such as microlenses, that are robust to perturbations like self-focusing. The network operates significantly faster than traditional CPU-based numerical solvers, offering a thousand-fold acceleration in computation times, making it suitable for inverse design and scattering applications. By integrating physical loss functions within the PyTorch framework, MaxwellNet provides an efficient way to solve complex nonlinear scattering problems while enabling the backpropagation of results for design optimization. This approach holds significant promise for accelerating the modelling and design of nanophotonic devices. Furthermore, the ability to perform fast computations and evaluate large sets of designs opens up new avenues for optimizing device shapes and improve the performance of nanophotonic systems in real-world conditions.[102]

Quantum optical systems leverage the unique properties of light at the quantum level and play a crucial role in advancing quantum technologies. The complex dynamics governing these systems present significant challenges in their modelling and simulation. Recently, PINNs have emerged as a powerful tool for addressing these challenges. By combining the robustness of neural networks with the principles of quantum mechanics, quantum neural networks (QNNs) have been developed to simulate and model quantum optical systems. These architectures offer a versatile approach to understanding quantum dynamics and enable more efficient solutions for a range of quantum information processing tasks..[103]

Combining quantum computing and PINNs has given rise to hybrid quantum-classical approaches for modelling quantum systems. Both methods leverage quantum computational dynamics within the PINN framework to solve complex quantum mechanical problems more efficiently. Dehaghani *et. al.* introduces a hybrid quantum-classical framework that integrates quantum computational dynamics with PINNs to address optimal control problems in quantum systems. By employing a dynamic quantum circuit with Gaussian and non-Gaussian gates, the model leverages classical machine learning techniques and solve state transition challenges in two- and three-level quantum systems. This approach offers a robust methodology for optimizing quantum state manipulations and ensure high precision in control dynamics. Its versatility in handling complex quantum state transitions underscores its potential for modelling nonlinear optical phenomena in nanophotonics. Therefore, advancing applications in quantum computing, secure communications, and precision metrology.[104]

## 2.4. Advantages and limitations

While the integration of PINNs in nanophotonics has shown promising results, several challenges remain. The complexity of nanophotonic structures requires sophisticated network architectures capable of capturing intricate physical phenomena. Additionally, the training of PINNs necessitates careful balancing of data-driven and physics-based components to ensure convergence and accuracy. Future research may focus on developing adaptive PINNs with trainable loss weights to improve accuracy and exploring the integration of PINNs with other AI techniques to enhance modelling capabilities.[60, 105, 106]

# 3. Deep Neural Networks in Electromagnetics

## 3.1. Electromagnetics Overview

Electromagnetics (EM) is the branch of physics that studies the interaction between electric charges and magnetic fields. It focuses on the phenomena's associated with electric and magnetic fields, electromagnetic radiation, and their interrelations. This field is foundational to many modern technologies, including wireless communication, radar, and electrical engineering systems.

## 3.2. DNN Application in electromagnetics

Over the past decade, deep learning (DL) has made incredible advances in EM applications, thanks to advancements in powerful computing systems the availability of large datasets. DL has emerged as a critical tool for addressing complex challenges in EM such as computational electromagnetics, antenna design, beamforming, and intelligent reflective surfaces. This section explores the diverse applications of DL in EM, highlighting its role across various domains and key applications.

### 3.2.1. Forward Scattering Problems

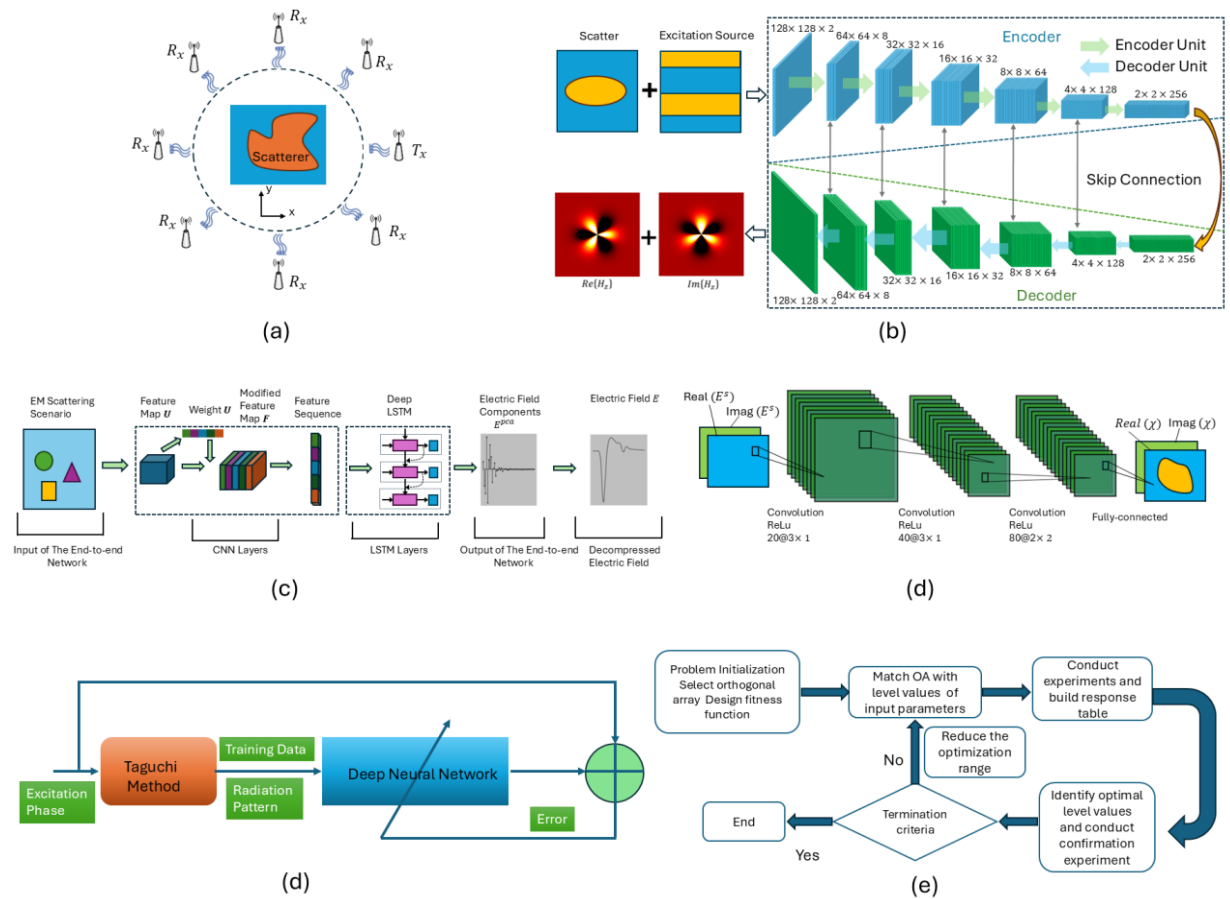

**Figure 6.** Illustration of various deep learning architectures applied to electromagnetic scattering problems. (a) Geometry of EM-scattering problems with 2-D TM. Here, Tx and Rx

represent the transmitting and receiving antennas, respectively, redrawn from.[107] (b) Input, output, and structure of the EM-net, redrawn from.[108] (c) Network architecture used for the prediction of electric field distributions, redrawn from.[109] (d) DConvNet architecture for the first step, redrawn from.[110] (e) Taguchi NNs training procedure, redrawn from.[111] (f) Flow chart of Taguchi's method, redrawn from.[111]

Forward Scattering Problems (FSPs) involve calculating the electromagnetic fields scattered by a known object when illuminated by an incident wave. Traditional computational methods, such as the finite-difference time-domain (FDTD) and finite element method (FEM), have long been foundational for solving FSPs in electromagnetics. These methods excel in accuracy and are widely used for modelling complex interactions governed by Maxwell's equations. However, they are computationally intensive, especially for large-scale or highly detailed simulations, which limit their feasibility for real-time applications [112]. This computational bottleneck is particularly significant as the demand for faster and more adaptable solutions increases in areas like 5G networks, radar systems, and real-time electromagnetic analysis.

DL algorithms, on the other hand, have shown to be a strong alternative to address these limitations, offering significantly faster solutions with comparable accuracy. Zhai et al. introduced a combination of CNNs and long short-term memory (LSTM) networks, which achieved a 1528-fold speedup over FDTD in predicting electromagnetic field distributions, with a very small average relative mean square error of approximately $8.05 \times 10^{-05}$ [109]. The network architecture is shown in Fig. 6a. Another approach was proposed by Qin et al., who utilized an LSTM-based framework that achieved a 113,924-fold speedup over FEM-TBC (an enhanced finite element method incorporating transitional boundary conditions) for predicting electromagnetic responses in hydraulic fractures [113]. Similarly, autoencoder-based networks have been utilized in FSPs in recent years for their ability to efficiently model complex electromagnetic interactions by learning detailed relationships between input conditions and output fields. Yao et al. applied a convolutional encoder-decoder network to electromagnetic scattering problems, effectively predicting total fields from incident waves and permittivity distributions [114]. Qi et al. proposed an autoencoder-based network approach to solve EM scattering problems involving complex geometries. Their method, referred to as EM-net, integrates residual architecture and skip connections to enhance convergence and generalization, as shown in Fig. 6b. This approach achieved high accuracy, with an average error of 1.23%, and a speedup of 2000 times compared to the traditional FDFD method [108].

Another promising approach involves employing FDTD-equivalent non-trainable neural networks. Bakirtzis et al. utilized graph neural networks (GNNs) [115], while Guo et al. implemented recurrent convolutional neural networks (RCNNs) [116]. These methods exploit the structural similarities between FDTD and neural network architectures, allowing them to bypass the need for training by directly determining network weights from the FDTD formulation. These advancements highlight the transformative potential of deep learning approaches in electromagnetic solvers, significantly reducing computation time while maintaining high accuracy and adaptability for real-time applications.

### 3.2.2. Inverse Scattering Problems

Inverse scattering problems (ISPs) aim to extract critical information, such as the location, shape, and material properties of an object, by analysing its scattered electromagnetic field. These problems are challenging due to their inherently ill-posed nature, where solutions may not be unique or stable, and their nonlinearity arising from multiple scattering effects [117]. The ISP setup is depicted in Fig. 6c. These problems require sophisticated techniques to extract meaningful information from scattered data, often involving iterative methods and regularization strategies to stabilize the solution process.

Neural networks (NNs) have been increasingly employed in ISPs to reconstruct the contrast of a target from scattered field measurements, showing promise in their ability to handle complex mappings and provide rapid, data-driven solutions that can outperform traditional iterative methods in speed and scalability. Ran et al. proposed a CNN for characterizing dielectric micro-structures from time-harmonic electromagnetic scattered fields [118]. The method was shown to outperform traditional contrast-source inversion (CSI) techniques, particularly for subwavelength structures and real-time reconstruction capabilities, achieving super-resolution imaging.

In another study, Fajardo et al. compared a Multilayer Perceptron (MLP) and a CNN for phaseless microwave imaging of dielectric cylinders. The CNN significantly outperformed the MLP in estimating both the geometric and dielectric properties of cylinders, especially in low-noise scenarios [119].

Similarly, Yao et al. introduced a two-step deep learning process for refining image reconstructions [110]. First, a complex-valued deep CNN retrieves the initial permittivity contrasts of dielectric scatterers from measured scattering data, as shown in Fig. 6d. These contrasts are then fed into a complex-valued deep residual CNN to refine the image reconstruction. While these approaches highlight the potential of deep learning for real-time imaging, they are generally more effective in reconstructing simpler scatterers and face challenges when dealing with more complex or highly nonlinear scenarios [107].

### 3.2.3. Antenna Design

The integration of DNNs into antenna design has gained considerable attention as the demand for wireless communication systems continues to grow, requiring higher data rates, improved performance, and more efficient designs. Traditional methods for antenna design, such as full-wave electromagnetic (EM) simulations and optimization algorithms, are typically resource-intensive and time-consuming, relying heavily on the experience of researchers and iterative trial-and-error processes. In contrast, DNNs' ability to model highly nonlinear relationships and handle large datasets makes them well-suited for predicting and optimizing antenna characteristics.

Several studies have demonstrated the advantages of employing DNNs for various antenna design and optimization tasks. Stanković *et al.* proposed a novel consensus deep neural network (C-DNN) to optimize the design of Yagi–Uda antennas.[120] The model was trained on datasets containing up to one million antenna samples generated via the Method of Moments

(MoM). The C-DNN significantly reduced design time by a factor of three to ten while maintaining accuracy within 0.5 dB of full-wave electromagnetic simulations, proving its effectiveness for optimizing antennas with complex parameter spaces.

Interesting contributions in antenna array design optimization were made by employing Taguchi-based neural networks, as described by Smida et al. [121]. Taguchi NNs uniquely integrate orthogonal array methodology with neural network learning capabilities to optimize excitation parameters, such as amplitude and phase, to achieve desired radiation patterns with low side-lobe levels and controlled null placements. Fig. 6e demonstrates the Taguchi NNs framework, while Fig. 6f illustrates the Taguchi method. Moreover, the incorporation of Radial Basis Function (RBF) networks into the Taguchi neural network framework has been shown by Oureghi et al. to enhance the handling of mutual coupling effects in antenna arrays [122]. RBF networks excel in nonlinear regression, enabling precise modelling of coupling effects and efficient optimization through Taguchi-generated datasets. This hybrid approach improves design accuracy and is ideal for adaptive, real-time applications like dynamic beam-steering, which is highly relevant for modern communication systems.

Another approach was proposed by Montaser et al. to optimize dual-band circularly polarized bone-shaped patch antennas operating at 28 GHz and 38 GHz for 5G wireless communication systems [123]. The neural network was trained using a hybrid algorithm combining the gravitational search algorithm and particle swarm optimization. The training utilized data from 150 simulated bone-shaped patch antennas, enabling accurate estimation of resonant frequencies. The study also applied deep learning for beam-steering in a 16-element circular antenna array, optimizing phase excitation to achieve side-lobe levels below -30 dB.

Yao et al. introduced a deep convolutional neural network (DConvNet) to detect unit failures in array antennas [124]. The DConvNet model was trained using far-field data generated from electromagnetic simulations, with Gaussian noise added to simulate real-world interference. This method demonstrated strong robustness to noise, achieving fault diagnosis accuracy rates of up to 100% even with significant interference, such as a signal-to-noise ratio (SNR) as low as 5 dB. This approach eliminates the need for complex Green's function computations or exhaustive measurements, significantly simplifying the fault detection process.

DNNs have also shown effectiveness in predicting antenna characteristics. Khan et al. developed a DNN model to estimate the far-field radiation patterns and reflection coefficient ($S_{11}$) of patch antennas in real-time [125]. The model utilized near-field data as input and was trained on 856,800 samples generated from various antenna geometries, substrate materials, and operating frequencies. This approach achieved a mean absolute percentage error of only 0.27%. Similarly, Kim et al. trained a DNN on 377 samples to accurately predict gamma-matching parameters for a traditional inverted-F antenna using only the magnitude of ($S_{11}$) as input to the network [126].

### 3.2.4. Beamforming and Intelligent Reflective Surfaces

Beamforming and intelligent reflecting surfaces (IRSs) are two critical technologies shaping the future of wireless communication, particularly in massive Multiple Input, Multiple Output (MIMO) systems and 6G networks. Beamforming, a key process in aligning an antenna array's main lobe with the direction of arrival (DoA) of a signal, traditionally relies on iterative optimization methods. While effective, these methods often struggle in dynamic environments like 5G networks, where real-time adaptability is crucial due to their computational complexity in achieving optimal signal-to-interference-plus-noise ratio (SINR) and minimal side-lobe levels (SLL) [127].

In contrast, neural network (NN)-based beamformers have demonstrated superior adaptability and efficiency, outperforming traditional methods, especially when incorporating realistic antenna parameters such as non-isotropic radiation patterns and mutual coupling between elements [128]. Recently, CNNs have been used to estimate broadband DoA with high accuracy and reduced computation time by analysing spatial covariance matrices, making them ideal for real-time applications [129].

Similarly, other studies have explored neural network approaches for adaptive beamforming, phased array synthesis, and interference rejection. Zaharis et al. introduced a neural network-based method for adaptive beamforming using data generated by a modified invasive weed optimization algorithm [130]. This approach trained the neural network to construct an adaptive beamformer capable of dynamically steering the main lobe toward desired signals while nullifying interference. The proposed method was able to match the accuracy of traditional beamformers while significantly reducing computational time.

In another study, Lovato et al. applied a CNN to phased array synthesis, specifically for generating antenna phases that achieve complex two-dimensional radiation patterns in an 8×8 microstrip patch antenna array [131]. The CNN was trained on diverse beam configurations and demonstrated high accuracy in synthesizing patterns, even for configurations that were not part of the training dataset. Ramezanpour et al. presented a two-stage neural network architecture combining a CNN and a bidirectional long short-term memory (bi-LSTM) network for interference rejection in beamforming systems [132]. The CNN estimated interference vectors, while the bi-LSTM focused on accurately modelling and extracting the desired signals. This hybrid approach proved particularly effective in high-interference environments, outperforming traditional beamforming techniques such as minimum mean square error (MMSE) and minimum variance distortionless response (MVDR). Moreover, it showed resilience to mismatches in the estimated autocorrelation matrix, a common issue in practical deployments.

These studies collectively highlight the versatility of neural networks in addressing diverse beamforming challenges, which become increasingly important when integrated with IRSs to maximize their performance potential. IRSs offer a promising approach to enhancing wireless communication by enabling precise control over electromagnetic wave propagation [133]. IRSs consist of periodic structures with reconfigurable elements that adjust the phase of incident signals—and, in some cases, amplitude and polarization—to optimize wireless channels, significantly improving energy efficiency without consuming additional transmit power [134]. However, optimizing the combined operation of beamformers and IRS phase shifters is challenging due to the complexity of configuring numerous passive elements, the need for accurate channel state information (CSI), and the challenges of non-convex optimization [135], [136].

To address these challenges, deep learning (DL) has been applied in many studies, driving advancements in IRS-based systems through improved channel estimation, symbol detection, and phase shift optimization. Sheen et al. demonstrated the effectiveness of CNNs in estimating both direct and cascaded channels, reducing the complexity typically associated with traditional methods [137]. Additionally, Zhang et al. proposed a deep denoising neural network to enhance compressive channel estimation, particularly for millimeter-wave systems, decreasing training overhead and maintaining robustness across various signal-to-noise ratios [138]. Khan et al. utilized deep residual networks to optimize phase configuration for channel mapping and signal detection, showcasing adaptability in multiple applications [139]. Furthermore, Wang et al. employed deep reinforcement learning (DRL) for real-time coarse phase control and optimization of IRS configurations, effectively addressing the challenges of dynamic environments without relying on detailed sub-channel state information (CSI) [140].

### 3.3. PINNs Applications in Electromagnetics

In recent years, PINNs have gained attention as a practical tool for tackling complex problems in various scientific fields. By embedding physical laws directly into the structure of neural networks, PINNs reduce the need for large datasets and combine the strengths of physics-based methods with deep learning techniques, making them particularly valuable for solving challenging multiphysics problems. PINNs solve multiphysics problems by incorporating partial differential equations and boundary conditions into their loss function, providing a powerful framework for solving both forward and inverse problems [105]. They bridge the gap between traditional physics-based methods and purely data-driven models to address complex computational challenges. Among the many applications of PINNs in electromagnetics, two primary areas stand out, which are the focus of this section.

#### 3.3.1. PINNs for Electromagnetics Simulations

Recently, several studies have successfully applied PINNs in computational electromagnetics (CEM) by directly incorporating Maxwell's equations into neural network architectures, offering a robust alternative to conventional numerical solvers. Research has demonstrated the versatility of PINNs across various CEM applications, ranging from steady-state analyses [141] to transient and multiscale scenarios [142].

Gong et al. employed a PINN to solve 2-D magnetostatic fields in electromagnetic devices [143]. The proposed method, shown in Fig. 7a, was validated against FEM and demonstrated the ability to solve for both the magnetic field intensity and the magnetic vector potential simultaneously, without requiring the spatial derivatives of constitutive parameters, which often introduce instability. Similarly, Beltrán-Pulido et al. proposed a technique for solving parametric magnetostatic problems using a variational principle and a neural network to model the magnetic vector potential [144]. This method stands out for its ability to efficiently handle high-dimensional design parameters while reducing reliance on traditional finite element methods (FEM).

However, several limitations still exist in these approaches. To improve accuracy and computational efficiency, hybrid methods that integrate labeled data and transfer learning have been introduced, demonstrating significant performance enhancements [141].

Su et al. proposed Physics-Informed Graph Neural Networks (PIGNNs) for electromagnetic simulations, addressing key challenges of traditional PINNs, such as ensuring field continuity [145]. PIGNNs, shown in Fig. 7b, excel at handling complex geometries and unstructured meshes while achieving FEM-level accuracy.

Similarly, Qi et al. introduced a new approach for multiphysics simulations by combining the FDTD method with PINNs to tackle coupled electromagnetic-thermal problems [146]. In this method, the FDTD algorithm computes electromagnetic fields and dissipated power density, which are then fed into a Physics-Informed U-Net (PIUN) to solve the heat equation, as shown in Fig. 7c. By embedding physical laws directly into the network, the PIUN operates without requiring pre-generated ground truth data or additional thermal solvers. This approach enables the computation of steady-state temperature distributions in lossy, inhomogeneous media while maintaining strong generalization capabilities for other multiphysics applications.

### 3.3.2. PINNs for Inverse Scattering Problems

Inverse scattering problems (ISPs) are one of the key areas where PINNs have demonstrated significant utility. Unlike conventional numerical methods, PINNs provide an efficient and mesh-free approach for solving ISPs by embedding governing equations and boundary conditions into their loss functions, simplifying the modeling process [147].

Recently, Hu et al. proposed a novel two-step inverse scattering method that integrates PINNs with a distorted finite-difference-frequency-domain-based iterative method (DFIM) [148]. In the first step, DFIM calculates the total fields and initial permittivity distributions of the target domain, which are then used as prior information. In the second step, this information is incorporated into the PINN framework as part of the loss function, significantly enhancing network accuracy and performance. The architecture of the proposed technique is illustrated in Fig. 7d.

One widely used method for solving inverse scattering problems is the U-Net architecture [149], which refines initial low-resolution contrast estimates using advanced image restoration techniques. This ability to transform low-resolution inputs into high-resolution outputs makes the U-Net an excellent choice for integrating PINNs into inverse problems. By incorporating preliminary estimates derived from non-iterative inversion methods—such as those obtained from back-propagation [150] and Born Approximation [151]—the network processes this information to deliver more accurate and computationally efficient solutions.

Xiao et al. employed a three-dimensional U-Net, shown in Fig. 7e, for electromagnetic inverse scattering in layered media [151]. The Born Approximation generates preliminary 3D images, which are then refined using Monte Carlo simulations, providing high-quality inputs for U-Net training. This process enables the 3D U-Net to outperform traditional iterative methods, such as the Variational Born Iteration Method, particularly in handling complex geometries and nonlinear scenarios.

Another promising approach is the SwitchNet architecture, introduced by Khoo et al., which utilizes sparse connections inspired by the Born Approximation to reduce the number of parameters required and facilitate training [152]. These advancements highlight how combining domain knowledge with deep learning architectures can significantly improve the accuracy and efficiency of inverse scattering solutions, unlike purely data-driven deep learning methods, which often struggle to handle highly nonlinear and complex scenarios.

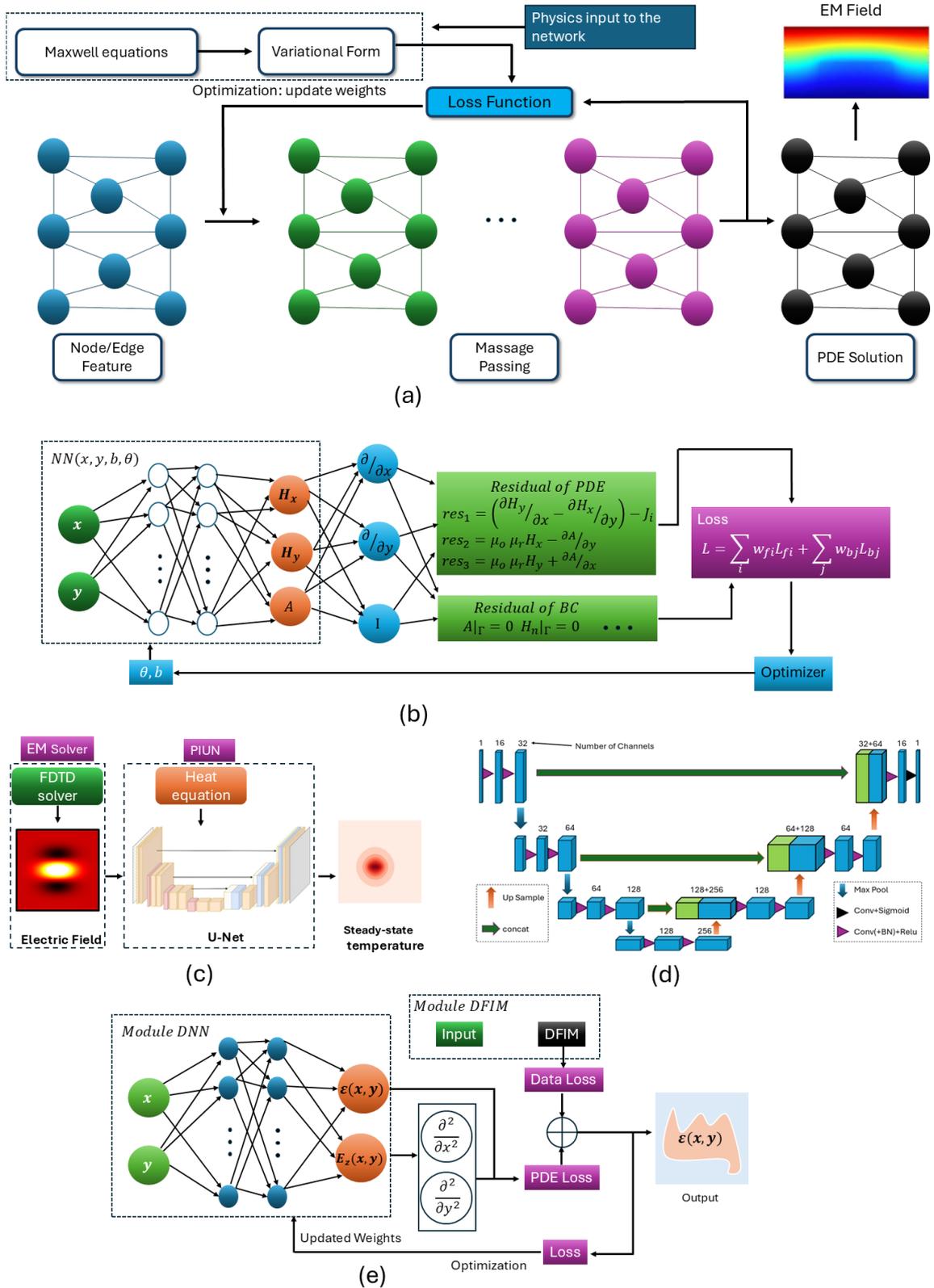

**Figure 7.** Illustration of various Physics informed neural network architectures applied to electromagnetic problems. (a) Schematic of the proposed PINN. θ, and b are the weights, and bias of NN, redrawn from.[143] (b) The general procedure to conduct PIGNN on a CEM problem, redrawn from.[145] (c) Multiphysics simulator: FDTD + physics-informed U-Net (PIUN), redrawn from.[146] (d) Schematic of the proposed DFIM-PINN, redrawn from.[148] (e) Architecture of the 3-D U-Net, redrawn from.[151]

## 3.4. Advantages and limitations

Neural networks offer notable advantages in electromagnetics, particularly in their ability to perform complex, data-intensive tasks such as antenna design, beamforming, and scattering analysis. These networks efficiently handle nonlinear and high-dimensional problems, significantly reducing computation time compared to traditional numerical methods such as FDTD and FEM [109]. Furthermore, traditional methods for tasks like beamforming often rely on iterative optimization techniques, which can be computationally intensive and slow, making them unsuitable for dynamic environments such as 5G networks. In contrast, deep learning approaches achieve solutions much faster, enabling real-time adaptability and responsiveness to meet the demands of modern applications. Deep learning models also often function as "black-box" solvers, because it difficult to interpret the output or incorporate domain-specific physics into their architectures. The lack of physical interpretability can lead to overfitting and reduced generalizability in unseen scenarios [107].

PINNs bridge these gaps by reducing reliance on extensive datasets while enhancing interpretability. PINNs naturally incorporate boundary conditions and physical principles to be effective for solving PDEs in electromagnetic problems. PINNs can achieve accurate results even in ill-posed or data-scarce environments by leveraging prior physical knowledge. They also demonstrate strong performance in multiscale problems, benefiting from fewer parameters and improved scalability compared to traditional deep learning methods [153].

Despite these strengths, PINNs face their own limitations. Training can be computationally intensive, especially for high-dimensional or transient problems [153]. Moreover, PINNs can struggle to capture high-frequency components due to spectral bias, leading to reduced accuracy when modelling sharp gradients or complex field variations [145]. The effectiveness of PINNs is also influenced by the quality of the embedded physical models, which may limit their applicability when these models are incomplete or imprecise.

Current trends in PINN research focus on addressing these challenges through hybrid frameworks, such as combining PINNs with graph neural networks [154] or leveraging advanced optimization techniques like the Non-Dominated Sorting Genetic Algorithm [155] to escape local minima and satisfy constraints.

# 4. Comparative Analysis and Benchmark

AI techniques have revolutionized nanophotonics and electromagnetic design by addressing challenges in modelling, optimization, and inverse design in several applications. We summarized and compared the key AI architectures used in these fields below and highlighted their strengths, limitations, and applications.

DNNs and CNNs architectures are widely employed for forward modelling and inverse design due to their ability to process high-dimensional data. Common applications range from light modulation to electromagnetic field distributions. CNNs excel in image-based tasks like metasurface geometry optimization, near field design, and microwave imaging, their reliance on large datasets and computational resources limits scalability. In contrast, PINNs enabling accurate inverse design with smaller datasets size. PINNs are particularly effective for wave propagation, resonant structure design, and modelling of electromagnetic problems. RL architecture dominated in inverse design of antenna design and resonance engineering. GANs efficiently explore vast inverse design problems such as light modulation, meta-hologram, and dielectric resonator antennas. Bayesian Optimization (BO) and Support Vector Machines (SVMs) provided lightweight alternatives for low-dimensional parameter tuning problems. However, they struggled with scalability of big model parameters. Emerging trends include hybrid models (e.g., PINN-GANs or PINN-CNNs) that combine physics-based constraints with generative capabilities for efficient computational usage. The current challenges remain in balancing computational efficiency, interpretability, and generalizability across different frequencies or applications.

**Table 1.** Summary of AI Architectures in Nanophotonics and EM Design

| AI Architecture | Design Type | PINN | Model Complexity (Params) | Application | Refs |
|---|---|---|---|---|---|
| DNNs | Forward Inverse | No | 1M–20M | Light phase engineering<br>Near field design<br>Light modulation<br>Cloaking<br>Resonance engineering<br>EMW scattering | 1, 3, 4, 7, 70, 72, 75, 152, 156-159 |
| CNNs | Forward Inverse | No | 1M–10M | Light modulation<br>Resonance engineering<br>Light phase engineering<br>Microwave Imaging | 2, 6, 73, 156, 160-162 |
| GAN | Inverse | No | 5M–20M | Light modulation<br>Meta hologram<br>Photonic crystal generation<br>Dielectric resonator antennas | 73, 74, 163-166 |
| VAE | Forward Inverse | No | 100k-5M | Light modulation<br>Resonance design | 80, 163, 167, 168 |

| PINN | Forward Inverse | Yes | 100k–1M | EMW propagation<br>Light time dynamics<br>Antenna design<br>EMW modelling | 11-15, 169 |
| RL | Inverse | No | 1M–5M | Antenna optimization<br>Programmable metasurface<br>Antenna Array Decoupling | 170-173 |
| BO | Inverse | No | <10k | Narrowband filter tuning<br>Resonance engineering | 174, 175 |
| SVM | Forward | No | <1k | EM scattering classification<br>Antenna Array Processing | 159, 176 |
| Transformer | Inverse | No | 10M–100M | Multi-scale EM design<br>Light modulation | 177, 178 |

## 5. Conclusion and Outlook

The recent advancements in AI, DNNs, and PINNs have revolutionized electromagnetic and nanophotonic design and applications. Key advantages include rapid design optimization, accurate modelling of complex light-matter interactions, and enhance design precision in applications such as meta-optics, quantum optics, and antenna design. PINNs stand out for their ability to unify physics-based constraints with data-driven learning. PINNs are superior in addressing both forward modelling and inverse design challenges while maintaining physical consistency with small size of dataset.

The integration of physical laws into PINNs is critical challenge as it introduces stiff, nonlinear loss landscapes that hinder convergence, particularly for multi-scale or high-dimensional systems. Balancing physics and data loss terms remains a persistent trade-off, which requires problem-specific tuning to avoid dominance of one loss over the other. Furthermore, the computational cost of convergence and evaluating physics losses across vast collocation points limits scalability for large-scale or geometrically complex systems. These bottlenecks require novel solutions such as adaptive loss-balancing strategies, hybrid models that synergize data efficiency with physics rigor, and algorithmic advances in sampling or parallelization to reduce training overhead.

According to our perspective, three key priorities will define progress in this field. First, large system optimization will push the development of adaptive PINN architectures through incorporating dynamic loss weighting, domain decomposition, and multi-fidelity modelling. Second, real-time optimization for industrial applications in telecommunications and quantum technologies will need bridging PINNs with emerging high speed computational platforms such as quantum machine learning or neuromorphic computing. Third, technology transfer of PINNs in experimental validation and predictive maintenance will foster the interdisciplinary collaboration and partnerships between academia and industry. Integrating science laws, modelling program, and experimental validation researchers will advance the design of novel physical and chemical devices with unprecedented functionality. A necessary step for developing innovations in sensing, chemical engineering, energy, and quantum information processing.

# References

(1) Yeung, C.; Tsai, J.-M.; King, B.; Pham, B.; Ho, D.; Liang, J.; Knight, M. W.; Raman, A. P. Multiplexed supercell metasurface design and optimization with tandem residual networks. *Nanophotonics* **2021**, *10* (3), 1133-1143.
(2) Sajedian, I.; Kim, J.; Rho, J. Finding the optical properties of plasmonic structures by image processing using a combination of convolutional neural networks and recurrent neural networks. *Microsystems & nanoengineering* **2019**, *5* (1), 27.
(3) An, S.; Fowler, C.; Zheng, B.; Shalaginov, M. Y.; Tang, H.; Li, H.; Zhou, L.; Ding, J.; Agarwal, A. M.; Rivero-Baleine, C. A deep learning approach for objective-driven all-dielectric metasurface design. *Acs Photonics* **2019**, *6* (12), 3196-3207.
(4) Peurifoy, J.; Shen, Y.; Jing, L.; Yang, Y.; Cano-Renteria, F.; DeLacy, B. G.; Joannopoulos, J. D.; Tegmark, M.; Soljačić, M. Nanophotonic particle simulation and inverse design using artificial neural networks. *Science advances* **2018**, *4* (6), eaar4206.
(5) Zhang, Z.; Lin, C.; Wang, B. Physics-informed shape optimization using coordinate projection. *Scientific Reports* **2024**, *14* (1), 6537.
(6) Wiecha, P. R.; Muskens, O. L. Deep learning meets nanophotonics: a generalized accurate predictor for near fields and far fields of arbitrary 3D nanostructures. *Nano letters* **2019**, *20* (1), 329-338.
(7) Zhen, Z.; Qian, C.; Jia, Y.; Fan, Z.; Hao, R.; Cai, T.; Zheng, B.; Chen, H.; Li, E. Realizing transmitted metasurface cloak by a tandem neural network. *Photonics Research* **2021**, *9* (5), B229-B235.
(8) Chen, M. K.; Liu, X.; Sun, Y.; Tsai, D. P. Artificial intelligence in meta-optics. *Chemical Reviews* **2022**, *122* (19), 15356-15413.
(9) Qian, C.; Zheng, B.; Shen, Y.; Jing, L.; Li, E.; Shen, L.; Chen, H. Deep-learning-enabled self-adaptive microwave cloak without human intervention. *Nature photonics* **2020**, *14* (6), 383-390.
(10) Antonelo, E. A.; Camponogara, E.; Seman, L. O.; Jordanou, J. P.; de Souza, E. R.; Hübner, J. F. Physics-informed neural nets for control of dynamical systems. *Neurocomputing* **2024**, *579*, 127419.
(11) Chen, Y.; Lu, L.; Karniadakis, G. E.; Dal Negro, L. Physics-informed neural networks for inverse problems in nano-optics and metamaterials. *Optics express* **2020**, *28* (8), 11618-11633.
(12) Tang, Y.; Fan, J.; Li, X.; Ma, J.; Qi, M.; Yu, C.; Gao, W. Physics-informed recurrent neural network for time dynamics in optical resonances. *Nature computational science* **2022**, *2* (3), 169-178.
(13) Liu, Y.-H.; Wang, B.-Z.; Wang, R. Inverse Design of Frequency Selective Surface Using Physics-Informed Neural Networks. *arXiv preprint arXiv:2401.03686* **2024**.

(14) Xu, Y.; Yang, J.-Q.; Fan, K.; Wang, S.; Wu, J.; Zhang, C.; Zhan, D.-C.; Padilla, W. J.; Jin, B.; Chen, J. Physics-informed Inverse Design of Multi-bit Programmable Metasurfaces. *arXiv preprint arXiv:2405.16795* **2024**.

(15) Medvedev, V.; Erdmann, A.; Rosskopf, A. Physics-informed deep learning for 3D modeling of light diffraction from optical metasurfaces. *Optics Express* **2025**, *33* (1), 1371-1384.

(16) Schmidhuber, J. Deep learning in neural networks: An overview. *Neural networks* **2015**, *61*, 85-117.

(17) Sonoda, S.; Murata, N. Neural network with unbounded activation functions is universal approximator. *Applied and Computational Harmonic Analysis* **2017**, *43* (2), 233-268.

(18) Soydaner, D. A comparison of optimization algorithms for deep learning. *International Journal of Pattern Recognition and Artificial Intelligence* **2020**, *34* (13), 2052013.

(19) Janocha, K.; Czarnecki, W. M. On loss functions for deep neural networks in classification. *arXiv preprint arXiv:1702.05659* **2017**.

(20) Gu, J.; Wang, Z.; Kuen, J.; Ma, L.; Shahroudy, A.; Shuai, B.; Liu, T.; Wang, X.; Wang, G.; Cai, J. Recent advances in convolutional neural networks. *Pattern recognition* **2018**, *77*, 354-377.

(21) Hochreiter, S.; Schmidhuber, J. Long short-term memory. *Neural computation* **1997**, *9* (8), 1735-1780.

(22) Berahmand, K.; Daneshfar, F.; Salehi, E. S.; Li, Y.; Xu, Y. Autoencoders and their applications in machine learning: a survey. *Artificial Intelligence Review* **2024**, *57* (2), 28.

(23) Ryu, I.; Park, G.-B.; Lee, Y.; Choi, D.-H. Physics-informed neural network for engineers: a review from an implementation aspect. *Journal of Mechanical Science and Technology* **2024**, *38* (7), 3499-3519.

(24) Voytan, D.; Sen, M. K. Wave propagation with physics informed neural networks. In *SEG International Exposition and Annual Meeting*, 2020; SEG: p D031S068R004.

(25) Liu, H.; Fan, Y.; Ding, F.; Du, L.; Zhao, J.; Sun, C.; Zhou, H. Physics-Informed Deep Model for Fast Time Domain Electromagnetic Simulation and Inversion. *IEEE Transactions on Antennas and Propagation* **2024**.

(26) Huang, B.; Wang, J. Applications of physics-informed neural networks in power systems-a review. *IEEE Transactions on Power Systems* **2022**, *38* (1), 572-588.

(27) Karniadakis, G. E.; Kevrekidis, I. G.; Lu, L.; Perdikaris, P.; Wang, S.; Yang, L. Physics-informed machine learning. *Nature Reviews Physics* **2021**, *3* (6), 422-440.

(28) Wu, Y.; Sicard, B.; Gadsden, S. A. Physics-informed machine learning: A comprehensive review on applications in anomaly detection and condition monitoring. *Expert Systems with Applications* **2024**, 124678.
(29) Jakšić, Z. Synergy between AI and Optical Metasurfaces: A Critical Overview of Recent Advances. In *Photonics*, 2024; MDPI: Vol. 11, p 442.
(30) Gao, H.; Sun, L.; Wang, J.-X. PhyGeoNet: Physics-informed geometry-adaptive convolutional neural networks for solving parameterized steady-state PDEs on irregular domain. *Journal of Computational Physics* **2021**, *428*, 110079.
(31) Zideh, M. J.; Solanki, S. K. Multivariate Physics-Informed Convolutional Autoencoder for Anomaly Detection in Power Distribution Systems with High Penetration of DERs. *arXiv preprint arXiv:2406.02927* **2024**.
(32) Zhang, R.; Liu, Y.; Sun, H. Physics-informed multi-LSTM networks for metamodeling of nonlinear structures. *Computer Methods in Applied Mechanics and Engineering* **2020**, *369*, 113226.
(33) Liu, F.; Li, J.; Wang, L. PI-LSTM: Physics-informed long short-term memory network for structural response modeling. *Engineering Structures* **2023**, *292*, 116500.
(34) Zheludev, N. I.; Kivshar, Y. S. From metamaterials to metadevices. *Nature materials* **2012**, *11* (11), 917-924.
(35) Jahani, S.; Jacob, Z. All-dielectric metamaterials. *Nature nanotechnology* **2016**, *11* (1), 23-36.
(36) Kildishev, A. V.; Boltasseva, A.; Shalaev, V. M. Planar photonics with metasurfaces. *Science* **2013**, *339* (6125), 1232009.
(37) Abdelraouf, O. A.; Wang, Z.; Liu, H.; Dong, Z.; Wang, Q.; Ye, M.; Wang, X. R.; Wang, Q. J.; Liu, H. Recent advances in tunable metasurfaces: materials, design, and applications. *ACS nano* **2022**, *16* (9), 13339-13369.
(38) Limonov, M. F.; Rybin, M. V.; Poddubny, A. N.; Kivshar, Y. S. Fano resonances in photonics. *Nature photonics* **2017**, *11* (9), 543-554.
(39) Kuznetsov, A. I.; Miroshnichenko, A. E.; Brongersma, M. L.; Kivshar, Y. S.; Luk'yanchuk, B. Optically resonant dielectric nanostructures. *Science* **2016**, *354* (6314), aag2472.
(40) Abdelraouf, O. A.; Allam, N. K. Towards nanostructured perovskite solar cells with enhanced efficiency: Coupled optical and electrical modeling. *Solar Energy* **2016**, *137*, 364-370.
(41) Abdelraouf, O. A.; Allam, N. K. Nanostructuring for enhanced absorption and carrier collection in CZTS-based solar cells: coupled optical and electrical modeling. *Optical Materials* **2016**, *54*, 84-88.
(42) Abdelraouf, O. A.; Abdelrahaman, M. I.; Allam, N. K. Plasmonic scattering nanostructures for efficient light trapping in flat czts solar cells. In *Metamaterials XI*, 2017; SPIE: Vol. 10227, pp 90-98.

(43) Abdelraouf, O. A.; Ali, H. A.; Allam, N. K. Optimizing absorption and scattering cross section of metal nanostructures for enhancing light coupling inside perovskite solar cells. In *2017 Conference on Lasers and Electro-Optics Europe & European Quantum Electronics Conference (CLEO/Europe-EQEC)*, 2017; IEEE: pp 1-1.
(44) Abdelraouf, O. A.; Shaker, A.; Allam, N. K. Front dielectric and back plasmonic wire grating for efficient light trapping in perovskite solar cells. *Optical materials* **2018**, *86*, 311-317.
(45) Abdelraouf, O. A.; Shaker, A.; Allam, N. K. Novel design of plasmonic and dielectric antireflection coatings to enhance the efficiency of perovskite solar cells. *Solar Energy* **2018**, *174*, 803-814.
(46) Abdelraouf, O. A.; Shaker, A.; Allam, N. K. Plasmonic nanoscatter antireflective coating for efficient CZTS solar cells. In *Photonics for Solar Energy Systems VII*, 2018; SPIE: Vol. 10688, pp 15-23.
(47) Abdelraouf, O. A.; Shaker, A.; Allam, N. K. Design of optimum back contact plasmonic nanostructures for enhancing light coupling in CZTS solar cells. In *Photonics for Solar Energy Systems VII*, 2018; SPIE: Vol. 10688, pp 33-41.
(48) Abdelraouf, O. A.; Shaker, A.; Allam, N. K. Design methodology for selecting optimum plasmonic scattering nanostructures inside CZTS solar cells. In *Photonics for Solar Energy Systems VII*, 2018; SPIE: Vol. 10688, pp 24-32.
(49) Abdelraouf, O. A.; Shaker, A.; Allam, N. K. Enhancing light absorption inside CZTS solar cells using plasmonic and dielectric wire grating metasurface. In *Metamaterials XI*, 2018; SPIE: Vol. 10671, pp 165-174.
(50) Abdelraouf, O. A.; Shaker, A.; Allam, N. K. All dielectric and plasmonic cross-grating metasurface for efficient perovskite solar cells. In *Metamaterials Xi*, 2018; SPIE: Vol. 10671, pp 104-112.
(51) Abdelraouf, O. A.; Shaker, A.; Allam, N. K. Using all dielectric and plasmonic cross grating metasurface for enhancing efficiency of CZTS solar cells. In *Nanophotonics VII*, 2018; SPIE: Vol. 10672, pp 246-255.
(52) Atef, N.; Emara, S. S.; Eissa, D. S.; El-Sayed, A.; Abdelraouf, O. A.; Allam, N. K. Well-dispersed Au nanoparticles prepared via magnetron sputtering on $TiO_2$ nanotubes with unprecedentedly high activity for water splitting. *Electrochemical Science Advances* **2021**, *1* (1), e2000004.
(53) Abdelraouf, O. A.; Anthur, A. P.; Dong, Z.; Liu, H.; Wang, Q.; Krivitsky, L.; Renshaw Wang, X.; Wang, Q. J.; Liu, H. Multistate tuning of third harmonic generation in fano-resonant hybrid dielectric metasurfaces. *Advanced Functional Materials* **2021**, *31* (48), 2104627.
(54) Abdelraouf, O. A.; Anthur, A. P.; Liu, H.; Dong, Z.; Wang, Q.; Krivitsky, L.; Wang, X. R.; Wang, Q. J.; Liu, H. Tunable transmissive THG in silicon metasurface enabled by phase change material. In *CLEO: QELS_Fundamental Science*, 2021; Optica Publishing Group: p FTh4K. 3.

(55) Abdelraouf, O. A.; Anthur, A. P.; Wang, X. R.; Wang, Q. J.; Liu, H. Modal phase-matched bound states in the continuum for enhancing third harmonic generation of deep ultraviolet emission. *ACS nano* **2024**, *18* (5), 4388-4397.

(56) Liu, H.; Wang, H.; Wang, H.; Deng, J.; Ruan, Q.; Zhang, W.; Abdelraouf, O. A.; Ang, N. S. S.; Dong, Z.; Yang, J. K. High-order photonic cavity modes enabled 3D structural colors. *ACS nano* **2022**, *16* (5), 8244-8252.

(57) Abdelraouf, O. A.; Wang, X. C.; Goh Ken, C. H.; Lim Nelson, C. B.; Ng, S. K.; Wang, W. D.; Renshaw Wang, X.; Wang, Q. J.; Liu, H. All-Optical Switching of Structural Color with a Fabry–Pérot Cavity. *Advanced Photonics Research* **2023**, *4* (11), 2300209.

(58) Jana, S.; Sreekanth, K. V.; Abdelraouf, O. A.; Lin, R.; Liu, H.; Teng, J.; Singh, R. Aperiodic Bragg reflectors for tunable high-purity structural color based on phase change material. *Nano Letters* **2024**, *24* (13), 3922-3929.

(59) Abdelraouf, O. A.; Wu, M.; Liu, H. Hybrid Metasurfaces Enabling Focused Tunable Amplified Photoluminescence Through Dual Bound States in the Continuum. *Advanced Functional Materials* **2025**, 2505165.

(60) Ji, W.; Chang, J.; Xu, H.-X.; Gao, J. R.; Gröblacher, S.; Urbach, H. P.; Adam, A. J. Recent advances in metasurface design and quantum optics applications with machine learning, physics-informed neural networks, and topology optimization methods. *Light: Science & Applications* **2023**, *12* (1), 169.

(61) Taflove, A.; Hagness, S. C.; Piket-May, M. Computational electromagnetics: the finite-difference time-domain method. *The Electrical Engineering Handbook* **2005**, *3* (629-670), 15.

(62) Jin, J.-M. *The finite element method in electromagnetics*; John Wiley & Sons, 2015.

(63) Oskooi, A. F.; Roundy, D.; Ibanescu, M.; Bermel, P.; Joannopoulos, J. D.; Johnson, S. G. MEEP: A flexible free-software package for electromagnetic simulations by the FDTD method. *Computer Physics Communications* **2010**, *181* (3), 687-702.

(64) Grosges, T.; Vial, A.; Barchiesi, D. Models of near-field spectroscopic studies: comparison between finite-element and finite-difference methods. *Optics Express* **2005**, *13* (21), 8483-8497.

(65) Lalanne, P.; Yan, W.; Vynck, K.; Sauvan, C.; Hugonin, J. P. Light interaction with photonic and plasmonic resonances. *Laser & Photonics Reviews* **2018**, *12* (5), 1700113.

(66) Ma, W.; Liu, Z.; Kudyshev, Z. A.; Boltasseva, A.; Cai, W.; Liu, Y. Deep learning for the design of photonic structures. *Nature Photonics* **2021**, *15* (2), 77-90.

(67) Khatib, O.; Ren, S.; Malof, J.; Padilla, W. J. Deep learning the electromagnetic properties of metamaterials—a comprehensive review. *Advanced Functional Materials* **2021**, *31* (31), 2101748.
(68) Wiecha, P. R.; Arbouet, A.; Girard, C.; Muskens, O. L. Deep learning in nano-photonics: inverse design and beyond. *Photonics Research* **2021**, *9* (5), B182-B200.
(69) Abdelraouf, O. A.; Mousa, A.; Ragab, M. NanoPhotoNet: AI-Enhanced Design Tool for Reconfigurable and High-Performance Multi-Layer Metasurfaces. *Photonics and Nanostructures-Fundamentals and Applications* **2025**, 101379.
(70) So, S.; Mun, J.; Rho, J. Simultaneous inverse design of materials and structures via deep learning: demonstration of dipole resonance engineering using core–shell nanoparticles. *ACS applied materials & interfaces* **2019**, *11* (27), 24264-24268.
(71) Tahersima, M. H.; Kojima, K.; Koike-Akino, T.; Jha, D.; Wang, B.; Lin, C.; Parsons, K. Deep neural network inverse design of integrated photonic power splitters. *Scientific reports* **2019**, *9* (1), 1368.
(72) Liu, D.; Tan, Y.; Khoram, E.; Yu, Z. Training deep neural networks for the inverse design of nanophotonic structures. *Acs Photonics* **2018**, *5* (4), 1365-1369.
(73) Liu, Z.; Zhu, D.; Rodrigues, S. P.; Lee, K.-T.; Cai, W. Generative model for the inverse design of metasurfaces. *Nano letters* **2018**, *18* (10), 6570-6576.
(74) Jiang, J.; Sell, D.; Hoyer, S.; Hickey, J.; Yang, J.; Fan, J. A. Free-form diffractive metagrating design based on generative adversarial networks. *ACS nano* **2019**, *13* (8), 8872-8878.
(75) Malkiel, I.; Mrejen, M.; Nagler, A.; Arieli, U.; Wolf, L.; Suchowski, H. Plasmonic nanostructure design and characterization via deep learning. *Light: Science & Applications* **2018**, *7* (1), 60.
(76) Asano, T.; Noda, S. Optimization of photonic crystal nanocavities based on deep learning. *Optics express* **2018**, *26* (25), 32704-32717.
(77) Jin, Y.; He, L.; Wen, Z.; Mortazavi, B.; Guo, H.; Torrent, D.; Djafari-Rouhani, B.; Rabczuk, T.; Zhuang, X.; Li, Y. Intelligent on-demand design of phononic metamaterials. *Nanophotonics* **2022**, *11* (3), 439-460.
(78) Tan, S.; Yang, F.; Boominathan, V.; Veeraraghavan, A.; Naik, G. V. 3D imaging using extreme dispersion in optical metasurfaces. *ACS Photonics* **2021**, *8* (5), 1421-1429.
(79) Gao, F.; Zhang, Z.; Xu, Y.; Zhang, L.; Yan, R.; Chen, X. Deep-learning-assisted designing chiral terahertz metamaterials with asymmetric transmission properties. *JOSA B* **2022**, *39* (6), 1511-1519.
(80) Ma, W.; Liu, Y. A data-efficient self-supervised deep learning model for design and characterization of nanophotonic structures. *Science China Physics, Mechanics & Astronomy* **2020**, *63* (8), 284212.

(81) Cheng, L.; Singh, P.; Ferranti, F. Transfer learning-assisted inverse modeling in nanophotonics based on mixture density networks. *IEEE Access* **2024**.
(82) Adibnia, E.; Mansouri-Birjandi, M. A.; Ghadrdan, M.; Jafari, P. A deep learning method for empirical spectral prediction and inverse design of all-optical nonlinear plasmonic ring resonator switches. *Scientific Reports* **2024**, *14* (1), 5787.
(83) Gao, L.; Qu, Y.; Wang, L.; Yu, Z. Computational spectrometers enabled by nanophotonics and deep learning. *Nanophotonics* **2022**, *11* (11), 2507-2529.
(84) Kojima, K.; Tahersima, M. H.; Koike-Akino, T.; Jha, D. K.; Tang, Y.; Wang, Y.; Parsons, K. Deep neural networks for inverse design of nanophotonic devices. *Journal of Lightwave Technology* **2021**, *39* (4), 1010-1019.
(85) Zhu, L.; Li, Y.; Yang, Z.; Zong, D.; Liu, Y. An On-Demand Inverse Design Method for Nanophotonic Devices Based on Generative Model and Hybrid Optimization Algorithm. *Plasmonics* **2024**, *19* (3), 1279-1290.
(86) Yan, R.; Wang, T.; Jiang, X.; Huang, X.; Wang, L.; Yue, X.; Wang, H.; Wang, Y. Efficient inverse design and spectrum prediction for nanophotonic devices based on deep recurrent neural networks. *Nanotechnology* **2021**, *32* (33), 335201.
(87) Frising, M.; Bravo-Abad, J.; Prins, F. Tackling multimodal device distributions in inverse photonic design using invertible neural networks. *Machine Learning: Science and Technology* **2023**, *4* (2), 02LT02.
(88) Li, Z.; Pestourie, R.; Lin, Z.; Johnson, S. G.; Capasso, F. Empowering metasurfaces with inverse design: principles and applications. *Acs Photonics* **2022**, *9* (7), 2178-2192.
(89) Liu, Z.; Zhu, D.; Raju, L.; Cai, W. Tackling photonic inverse design with machine learning. *Advanced Science* **2021**, *8* (5), 2002923.
(90) Nadell, C. C.; Huang, B.; Malof, J. M.; Padilla, W. J. Deep learning for accelerated all-dielectric metasurface design. *Optics express* **2019**, *27* (20), 27523-27535.
(91) Chen, Y.; Dal Negro, L. Physics-informed neural networks for imaging and parameter retrieval of photonic nanostructures from near-field data. *APL Photonics* **2022**, *7* (1).
(92) Khatib, O.; Ren, S.; Malof, J.; Padilla, W. J. Learning the physics of all-dielectric metamaterials with deep Lorentz neural networks. *Advanced Optical Materials* **2022**, *10* (13), 2200097.
(93) Liang, B.; Xu, D.; Yu, N.; Xu, Y.; Ma, X.; Liu, Q.; Asif, M. S.; Yan, R.; Liu, M. Physics-guided neural-network-based inverse design of a photonic–plasmonic nanodevice for superfocusing. *ACS Applied Materials & Interfaces* **2022**, *14* (23), 27397-27404.

(94) Wang, Y.; Zhang, S. Multi-receptive-field physics-informed neural network for complex electromagnetic media. *Optical Materials Express* **2024**, *14* (11), 2740-2754.
(95) Sui, X.; Wu, Q.; Liu, J.; Chen, Q.; Gu, G. A review of optical neural networks. *IEEE Access* **2020**, *8*, 70773-70783.
(96) Chen, M.; Lupoiu, R.; Mao, C.; Huang, D.-H.; Jiang, J.; Lalanne, P.; Fan, J. A. High speed simulation and freeform optimization of nanophotonic devices with physics-augmented deep learning. *ACS Photonics* **2022**, *9* (9), 3110-3123.
(97) Park, C.; Kim, S.; Jung, A. W.; Park, J.; Seo, D.; Kim, Y.; Park, C.; Park, C. Y.; Jang, M. S. Sample-efficient inverse design of freeform nanophotonic devices with physics-informed reinforcement learning. *Nanophotonics* **2024**, *13* (8), 1483-1492.
(98) Jiang, X.; Wang, D.; Fan, Q.; Zhang, M.; Lu, C.; Lau, A. P. T. Physics-Informed Neural Network for Nonlinear Dynamics in Fiber Optics. *Laser & Photonics Reviews* **2022**, *16* (9), 2100483.
(99) Linka, K.; Schäfer, A.; Meng, X.; Zou, Z.; Karniadakis, G. E.; Kuhl, E. Bayesian Physics Informed Neural Networks for real-world nonlinear dynamical systems. *Computer Methods in Applied Mechanics and Engineering* **2022**, *402*, 115346.
(100) Jiang, X.; Wang, D.; Chen, X.; Zhang, M. Physics-informed neural network for optical fiber parameter estimation from the nonlinear Schrödinger equation. *Journal of Lightwave Technology* **2022**, *40* (21), 7095-7105.
(101) Liu, D.; Zhang, W.; Gao, Y.; Fan, D.; Malomed, B. A.; Zhang, L. Physics-informed neural network for nonlinear dynamics of self-trapped necklace beams. *Optics Express* **2024**, *32* (22), 38531-38549.
(102) Gigli, C.; Saba, A.; Ayoub, A. B.; Psaltis, D. Predicting nonlinear optical scattering with physics-driven neural networks. *Apl Photonics* **2023**, *8* (2).
(103) Markidis, S. On physics-informed neural networks for quantum computers. *Frontiers in Applied Mathematics and Statistics* **2022**, *8*, 1036711.
(104) Dehaghani, N. B.; Aguiar, A. P.; Wisniewski, R. A Hybrid Quantum-Classical Physics-Informed Neural Network Architecture for Solving Quantum Optimal Control Problems. *arXiv preprint arXiv:2404.15015* **2024**.
(105) Raissi, M.; Perdikaris, P.; Karniadakis, G. E. Physics-informed neural networks: A deep learning framework for solving forward and inverse problems involving nonlinear partial differential equations. *Journal of Computational physics* **2019**, *378*, 686-707.
(106) Lawal, Z. K.; Yassin, H.; Lai, D. T. C.; Che Idris, A. Physics-informed neural network (PINN) evolution and beyond: A systematic literature review and bibliometric analysis. *Big Data and Cognitive Computing* **2022**, *6* (4), 140.

(107) Chen, X.; Wei, Z.; Maokun, L.; Rocca, P. A review of deep learning approaches for inverse scattering problems (invited review). *Electromagnetic Waves* **2020**, *167*, 67-81.

(108) Qi, S.; Wang, Y.; Li, Y.; Wu, X.; Ren, Q.; Ren, Y. Two-dimensional electromagnetic solver based on deep learning technique. *IEEE Journal on Multiscale and Multiphysics Computational Techniques* **2020**, *5*, 83-88.

(109) Zhai, M.; Chen, Y.; Xu, L.; Yin, W.-Y. An end-to-end neural network for complex electromagnetic simulations. *IEEE Antennas and Wireless Propagation Letters* **2023**.

(110) Yao, H. M.; Wei, E.; Jiang, L. Two-step enhanced deep learning approach for electromagnetic inverse scattering problems. *IEEE Antennas and Wireless Propagation Letters* **2019**, *18* (11), 2254-2258.

(111) Weng, W.-C.; Yang, F.; Elsherbeni, A. Z. Linear antenna array synthesis using Taguchi's method: A novel optimization technique in electromagnetics. *IEEE Transactions on Antennas and Propagation* **2007**, *55* (3), 723-730.

(112) Massa, A.; Marcantonio, D.; Chen, X.; Li, M.; Salucci, M. DNNs as applied to electromagnetics, antennas, and propagation—A review. *IEEE Antennas and Wireless Propagation Letters* **2019**, *18* (11), 2225-2229.

(113) Qin, H.; Qin, W.; Kang, Z.; Chen, Y. An efficient forward modeling method of electromagnetic response of multiscale hydraulic fracture based on deep learning. *IEEE Antennas and Wireless Propagation Letters* **2024**.

(114) Yao, H. M.; Jiang, L.; Ng, M. Implementing the Fast Full-Wave Electromagnetic Forward Solver Using the Deep Convolutional Encoder-Decoder Architecture. *IEEE Transactions on Antennas and Propagation* **2022**, *71* (1), 1152-1157.

(115) Bakirtzis, S.; Fiore, M.; Zhang, J.; Wassell, I. Solving Maxwell's equations with Non-Trainable Graph Neural Network Message Passing. *arXiv preprint arXiv:2405.00814* **2024**.

(116) Guo, L.; Li, M.; Xu, S.; Yang, F.; Liu, L. Electromagnetic modeling using an FDTD-equivalent recurrent convolution neural network: Accurate computing on a deep learning framework. *IEEE Antennas and Propagation Magazine* **2021**, *65* (1), 93-102.

(117) Chen, X. *Computational methods for electromagnetic inverse scattering*; John Wiley & Sons, 2018.

(118) Ran, P.; Qin, Y.; Lesselier, D. Electromagnetic imaging of a dielectric micro-structure via convolutional neural networks. In *2019 27th European Signal Processing Conference (EUSIPCO)*, 2019; IEEE: pp 1-5.

(119) Fajardo, J. E.; Galván, J.; Vericat, F.; Carlevaro, C. M.; Irastorza, R. M. Phaseless microwave imaging of dielectric cylinders: An artificial neural networks-based approach. *arXiv preprint arXiv:1908.10424* **2019**.

(120) Stanković, Z. Ž.; Olćan, D. I.; Dončov, N. S.; Kolundžija, B. M. Consensus deep neural networks for antenna design and optimization. *IEEE Transactions on Antennas and Propagation* **2021**, *70* (7), 5015-5023.
(121) Smida, A.; Ghayoula, R.; Nemri, N.; Trabelsi, H.; Gharsallah, A.; Grenier, D. Phased arrays in communication system based on Taguchi-neural networks. *International Journal of Communication Systems* **2014**, *27* (12), 4449-4466.
(122) Oureghi, K.; Ghayoula, R.; Amara, W.; Smida, A.; El Gmati, I.; Fattahi, J. Taguchi-rbf neural networks based optimization of phased array antenna with coupling effects. *Advanced Electromagnetics* **2023**, *12* (1), 35-44.
(123) Montaser, A. M.; Mahmoud, K. R. Deep learning based antenna design and beam-steering capabilities for millimeter-wave applications. *IEEE Access* **2021**, *9*, 145583-145591.
(124) Yao, H. M.; Li, M.; Jiang, L.; Yeung, K. L.; Ng, M. Antenna Array Diagnosis Using Deep Learning Approach. *IEEE Transactions on Antennas and Propagation* **2024**.
(125) Khan, M. R.; Zekios, C. L.; Bhardwaj, S.; Georgakopoulos, S. V. A generalized approach to real-time performance estimation of antenna types using deep learning. In *2022 IEEE International Symposium on Antennas and Propagation and USNC-URSI Radio Science Meeting (AP-S/URSI)*, 2022; IEEE: pp 497-498.
(126) Kim, J. H.; Bang, J. Antenna impedance matching using deep learning. *Sensors* **2021**, *21* (20), 6766.
(127) Ali, E.; Ismail, M.; Nordin, R.; Abdulah, N. F. Beamforming techniques for massive MIMO systems in 5G: overview, classification, and trends for future research. *Frontiers of Information Technology & Electronic Engineering* **2017**, *18*, 753-772.
(128) Zaharis, Z. D.; Gravas, I. P.; Lazaridis, P. I.; Yioultsis, T. V.; Antonopoulos, C. S.; Xenos, T. D. An effective modification of conventional beamforming methods suitable for realistic linear antenna arrays. *IEEE Transactions on Antennas and Propagation* **2020**, *68* (7), 5269-5279.
(129) Zhu, W.; Zhang, M. A deep learning architecture for broadband DOA estimation. In *2019 IEEE 19th International Conference on Communication Technology (ICCT)*, 2019; IEEE: pp 244-247.
(130) Zaharis, Z. D.; Skeberis, C.; Xenos, T. D.; Lazaridis, P. I.; Cosmas, J. Design of a novel antenna array beamformer using neural networks trained by modified adaptive dispersion invasive weed optimization based data. *IEEE Transactions on Broadcasting* **2013**, *59* (3), 455-460.
(131) Lovato, R.; Gong, X. Phased antenna array beamforming using convolutional neural networks. In *2019 IEEE International Symposium on Antennas and Propagation and USNC-URSI Radio Science Meeting*, 2019; IEEE: pp 1247-1248.

(132) Ramezanpour, P.; Mosavi, M.-R. Two-stage beamforming for rejecting interferences using deep neural networks. *IEEE Systems Journal* **2020**, *15* (3), 4439-4447.
(133) Zhao, J. A survey of intelligent reflecting surfaces (IRSs): Towards 6G wireless communication networks. *arXiv preprint arXiv:1907.04789* **2019**.
(134) Huang, C.; Zappone, A.; Alexandropoulos, G. C.; Debbah, M.; Yuen, C. Reconfigurable intelligent surfaces for energy efficiency in wireless communication. *IEEE transactions on wireless communications* **2019**, *18* (8), 4157-4170.
(135) Al Kassir, H.; Zaharis, Z. D.; Lazaridis, P. I.; Kantartzis, N. V.; Yioultsis, T. V.; Xenos, T. D. A review of the state of the art and future challenges of deep learning-based beamforming. *IEEE Access* **2022**, *10*, 80869-80882.
(136) Zhou, Z.; Bai, K.; Mohammadi, N.; Yi, Y.; Liu, L. Making intelligent reflecting surfaces more intelligent: A roadmap through reservoir computing. *IEEE Network* **2022**, *36* (2), 175-180.
(137) Sheen, B.; Yang, J.; Feng, X.; Chowdhury, M. M. U. A deep learning based modeling of reconfigurable intelligent surface assisted wireless communications for phase shift configuration. *IEEE Open Journal of the Communications Society* **2021**, *2*, 262-272.
(138) Zhang, X.; Li, G.; Zhang, J.; Hu, A.; Hou, Z.; Xiao, B. Deep-learning-based physical-layer secret key generation for FDD systems. *IEEE Internet of Things Journal* **2021**, *9* (8), 6081-6094.
(139) Khan, S.; Khan, K. S.; Haider, N.; Shin, S. Y. Deep-learning-aided detection for reconfigurable intelligent surfaces. *arXiv preprint arXiv:1910.09136* **2019**.
(140) Wang, W.; Zhang, W. Intelligent reflecting surface configurations for smart radio using deep reinforcement learning. *IEEE Journal on Selected Areas in Communications* **2022**, *40* (8), 2335-2346.
(141) Khan, A.; Lowther, D. A. Physics informed neural networks for electromagnetic analysis. *IEEE Transactions on Magnetics* **2022**, *58* (9), 1-4.
(142) Noakoasteen, O.; Wang, S.; Peng, Z.; Christodoulou, C. Physics-informed deep neural networks for transient electromagnetic analysis. *IEEE Open Journal of Antennas and Propagation* **2020**, *1*, 404-412.
(143) Gong, Z.; Chu, Y.; Yang, S. Physics-Informed Neural Networks for Solving 2-D Magnetostatic Fields. *IEEE Transactions on Magnetics* **2023**, *59* (11), 1-5.
(144) Beltrán-Pulido, A.; Bilionis, I.; Aliprantis, D. Physics-informed neural networks for solving parametric magnetostatic problems. *IEEE Transactions on Energy Conversion* **2022**, *37* (4), 2678-2689.
(145) Su, Y.; Zeng, S.; Wu, X.; Huang, Y.; Chen, J. Physics-Informed Graph Neural Network for Electromagnetic Simulations. In *2023 XXXVth General Assembly and*

*Scientific Symposium of the International Union of Radio Science (URSI GASS)*, 2023; IEEE: pp 1-3.
(146) Qi, S.; Sarris, C. D. Physics-Informed Neural Networks for Multiphysics Simulations: Application to Coupled Electromagnetic-Thermal Modeling. In *2023 IEEE/MTT-S International Microwave Symposium-IMS 2023*, 2023; IEEE: pp 166-169.
(147) Chen, X.-X.; Zhang, P.; Yin, Z.-Y. Physics-Informed neural network solver for numerical analysis in geoengineering. *Georisk: Assessment and Management of Risk for Engineered Systems and Geohazards* **2024**, *18* (1), 33-51.
(148) Hu, Y.-D.; Wang, X.-H.; Zhou, H.; Wang, L. A Priori Knowledge-Based Physics-Informed Neural Networks for Electromagnetic Inverse Scattering. *IEEE Transactions on Geoscience and Remote Sensing* **2024**.
(149) Ronneberger, O.; Fischer, P.; Brox, T. U-net: Convolutional networks for biomedical image segmentation. In *Medical image computing and computer-assisted intervention–MICCAI 2015: 18th international conference, Munich, Germany, October 5-9, 2015, proceedings, part III 18*, 2015; Springer: pp 234-241.
(150) Sun, Y.; Xia, Z.; Kamilov, U. S. Efficient and accurate inversion of multiple scattering with deep learning. *Optics express* **2018**, *26* (11), 14678-14688.
(151) Xiao, J.; Li, J.; Chen, Y.; Han, F.; Liu, Q. H. Fast electromagnetic inversion of inhomogeneous scatterers embedded in layered media by born approximation and 3-D U-Net. *IEEE Geoscience and Remote Sensing Letters* **2019**, *17* (10), 1677-1681.
(152) Khoo, Y.; Ying, L. SwitchNet: a neural network model for forward and inverse scattering problems. *SIAM Journal on Scientific Computing* **2019**, *41* (5), A3182-A3201.
(153) Liu, Z.; Xu, F. Principle and application of physics-inspired neural networks for electromagnetic problems. In *IGARSS 2022-2022 IEEE International Geoscience and Remote Sensing Symposium*, 2022; IEEE: pp 5244-5247.
(154) Gao, H.; Zahr, M. J.; Wang, J.-X. Physics-informed graph neural Galerkin networks: A unified framework for solving PDE-governed forward and inverse problems. *Computer Methods in Applied Mechanics and Engineering* **2022**, *390*, 114502.
(155) Lu, B.; Moya, C.; Lin, G. NSGA-PINN: a multi-objective optimization method for physics-informed neural network training. *Algorithms* **2023**, *16* (4), 194.
(156) Ma, W.; Cheng, F.; Liu, Y. Deep-learning-enabled on-demand design of chiral metamaterials. *ACS nano* **2018**, *12* (6), 6326-6334.
(157) So, S.; Yang, Y.; Lee, T.; Rho, J. On-demand design of spectrally sensitive multiband absorbers using an artificial neural network. *Photonics Research* **2021**, *9* (4), B153-B158.

(158) Zhang, J.; Wang, G.; Wang, T.; Li, F. Genetic algorithms to automate the design of metasurfaces for absorption bandwidth broadening. *ACS Applied Materials & Interfaces* **2021**, *13* (6), 7792-7800.

(159) Zhelyeznyakov, M. V.; Brunton, S.; Majumdar, A. Deep learning to accelerate scatterer-to-field mapping for inverse design of dielectric metasurfaces. *ACS Photonics* **2021**, *8* (2), 481-488.

(160) Elsawy, M. M.; Lanteri, S.; Duvigneau, R.; Fan, J. A.; Genevet, P. Numerical optimization methods for metasurfaces. *Laser & Photonics Reviews* **2020**, *14* (10), 1900445.

(161) An, S.; Zheng, B.; Shalaginov, M. Y.; Tang, H.; Li, H.; Zhou, L.; Ding, J.; Agarwal, A. M.; Rivero-Baleine, C.; Kang, M. Deep learning modeling approach for metasurfaces with high degrees of freedom. *Optics Express* **2020**, *28* (21), 31932-31942.

(162) Aydın, İ.; Budak, G.; Sefer, A.; Yapar, A. CNN-based deep learning architecture for electromagnetic imaging of rough surface profiles. *IEEE Transactions on Antennas and Propagation* **2022**, *70* (10), 9752-9763.

(163) Liu, C.; Yu, W. M.; Ma, Q.; Li, L.; Cui, T. J. Intelligent coding metasurface holograms by physics-assisted unsupervised generative adversarial network. *Photonics Research* **2021**, *9* (4), B159-B167.

(164) So, S.; Rho, J. Designing nanophotonic structures using conditional deep convolutional generative adversarial networks. *Nanophotonics* **2019**, *8* (7), 1255-1261.

(165) Wang, H. P.; Li, Y. B.; Li, H.; Dong, S. Y.; Liu, C.; Jin, S.; Cui, T. J. Deep learning designs of anisotropic metasurfaces in ultrawideband based on generative adversarial networks. *Advanced Intelligent Systems* **2020**, *2* (9), 2000068.

(166) Liu, M.; Zhang, H.; Song, J.; Lu, M. Using generative model for intelligent design of dielectric resonator antennas. *Microwave and Optical Technology Letters* **2024**, *66* (1), e34013.

(167) Liu, Z.; Raju, L.; Zhu, D.; Cai, W. A hybrid strategy for the discovery and design of photonic structures. *IEEE Journal on Emerging and Selected Topics in Circuits and Systems* **2020**, *10* (1), 126-135.

(168) Ma, W.; Cheng, F.; Xu, Y.; Wen, Q.; Liu, Y. Probabilistic representation and inverse design of metamaterials based on a deep generative model with semi-supervised learning strategy. *Advanced Materials* **2019**, *31* (35), 1901111.

(169) Wei, Z.; Zhou, Z.; Wang, P.; Ren, J.; Yin, Y.; Pedersen, G. F.; Shen, M. Fully automated design method based on reinforcement learning and surrogate modeling for antenna array decoupling. *IEEE Transactions on Antennas and Propagation* **2022**, *71* (1), 660-671.

(170) Li, R.; Zhang, C.; Xie, W.; Gong, Y.; Ding, F.; Dai, H.; Chen, Z.; Yin, F.; Zhang, Z. Deep reinforcement learning empowers automated inverse design and optimization of photonic crystals for nanoscale laser cavities. *Nanophotonics* **2023**, *12* (2), 319-334.

(171) Zhao, Y.; Li, L.; Lanteri, S.; Viquerat, J. Dynamic metasurface control using deep reinforcement learning. *Mathematics and Computers in Simulation* **2022**, *197*, 377-395.

(172) Hu, J.; Zhang, H.; Bian, K.; Di Renzo, M.; Han, Z.; Song, L. MetaSensing: Intelligent metasurface assisted RF 3D sensing by deep reinforcement learning. *IEEE Journal on Selected Areas in Communications* **2021**, *39* (7), 2182-2197.

(173) Nohra, M.; Dufour, S. Physics-Informed Neural Networks for the Numerical Modeling of Steady-State and Transient Electromagnetic Problems with Discontinuous Media. *arXiv preprint arXiv:2406.04380* **2024**.

(174) Wray, P. R.; Paul, E. G.; Atwater, H. A. Optical filters made from random metasurfaces using Bayesian optimization. *Nanophotonics* **2024**, *13* (2), 183-193.

(175) Zeng, Y.; Qing, X.; Chia, M. Y.-W. A wideband circularly polarized antenna with a non-uniform metasurface designed via multi-objective Bayesian optimization. *IEEE Antennas and Wireless Propagation Letters* **2024**.

(176) Xu, N.; Christodoulou, C.; Martinez-Ramon, M.; Ozdemir, T. Antenna array processing for radar applications using support vector machines. In *2006 IEEE Antennas and Propagation Society International Symposium*, 2006; IEEE: pp 1295-1298.

(177) Jin, L.; Huang, Y.-W.; Jin, Z.; Devlin, R. C.; Dong, Z.; Mei, S.; Jiang, M.; Chen, W. T.; Wei, Z.; Liu, H. Dielectric multi-momentum meta-transformer in the visible. *Nature communications* **2019**, *10* (1), 4789.

(178) Hu, Q.; Chen, K.; Zhao, J.; Dong, S.; Jiang, T.; Feng, Y. On-demand dynamic polarization meta-transformer. *Laser & Photonics Reviews* **2023**, *17* (1), 2200479.